\documentclass[review]{elsarticle}
\usepackage{setspace}
\usepackage{enumerate}
\usepackage{graphicx}
\usepackage{amsmath, bm}
\usepackage{float}
\usepackage{mdframed}
\usepackage{multirow}
\usepackage{textcomp}
\usepackage{subcaption}
\usepackage{tikz}
\usepackage[utf8]{inputenc}
\usepackage{newunicodechar}
\newunicodechar{−}{-}
\usepackage{float}
\usetikzlibrary{
arrows.meta,
positioning,
shapes.geometric,
fit,
calc
}

\usepackage{amsmath}
\usepackage{url}
\usepackage{makecell}
\usepackage{amssymb}
\usepackage{bm}
\usepackage{textcmds}
 \usepackage{hyperref}
\biboptions{numbers,sort&compress}
\usepackage{tikz}
\usetikzlibrary{arrows.meta, positioning, shapes.geometric}
\usetikzlibrary{calc}
\tikzstyle{nnblock} = [
    draw, dashed, rounded corners,
    minimum width=5cm, minimum height=4cm
]

\tikzstyle{physblock} = [
    draw, dashed, rounded corners,
    minimum width=5cm, minimum height=4cm
]

\tikzstyle{neuron} = [
    draw, circle, fill=orange!25,
    minimum size=8mm
]

\tikzstyle{param} = [
    draw, circle, fill=purple!25,
    minimum size=8mm
]

\tikzstyle{loss} = [
    draw, ellipse, fill=green!20,
    minimum width=2.8cm
]

\usepackage{pgfplots}
\pgfplotsset{compat=1.18}
\usepackage{array}

\usetikzlibrary{shapes.geometric, arrows}
\usepackage{caption}  

\def\be{\begin{equation}}\def\ee{\end{equation}}
\def\ba{\begin{array}}\def\ea{\end{array}}
\def\bfg{\begin{figure}}\def\efg{\end{figure}}
\makeatletter
\def\fps@figure{htbp}
\makeatother

\usepackage{stackengine}
\stackMath
\newcommand\tenq[2][1]{%
 \def\useanchorwidth{T}%
  \ifnum#1>1%
    \stackunder[0pt]{\tenq[\numexpr#1-1\relax]{#2}}{\scriptscriptstyle\sim}%
  \else%
    \stackunder[1pt]{#2}{\scriptscriptstyle\sim}%
  \fi%
}

\RequirePackage[margin=20mm]{geometry}

\journal{}
\begin{document}

\begin{frontmatter}

\title{Physics-Informed Neural Network Approach for Surface Wave Propagation in Functionally Graded Magnetoelastic Layered Media}

\author[label1]{Diksha}
\author[label1]{Katyayani}
\author[label1]{Hriticka Dhiman}
\author[label1]{Soniya Chaudhary*}
\author[label1]{Pawan Kumar Sharma}
\author[label2]{Mayank Kumar Jha}
\cortext[cor1]{Corresponding author: soniyachaudhary18@gmail.com}
\address[label1]{Department of Mathematics and Scientific Computing, National Institute of Technology Hamirpur, Himachal Pradesh, 177005, India}
\address[label2]{TEKsystems (HP), Banglore, Karnataka, 560035, India}

\begin{abstract}
This paper investigates propagation of SH-waves in a layered composite structure consisting of a pre-stressed functionally graded magnetoelastic orthotropic layer overlying a pre-stressed functionally graded orthotropic half-space under the influence of gravity. The study introduces a physics-informed neural network (PINN) framework for the dispersion analysis of SH-waves in the considered composite medium. As a benchmark, an analytical solution to the dispersion relation is derived and used to validate accuracy and reliability of the proposed PINN formulation.
In the developed PINN model, the phase velocity corresponding to a prescribed wave number is treated as a trainable parameter, enabling the determination of the dispersion relation associated with the nonlinear eigenvalue problem. The Adam optimizer is employed to minimize the loss function during the training process. In addition, the effects of different activation functions and network architectures, including variations in number of hidden layers and neurons, are systematically investigated to study the performance of the proposed framework. Error analysis is carried out using several norms, namely $L_1$, $L_2$, RMSE, relative absolute error, and $L_\infty$, to assess the accuracy of the predictions. Furthermore, the variation of phase velocity with wave number under different material parameters is investigated. The comparison between the analytical and PINN-based results demonstrates excellent agreement, confirming the effectiveness of the proposed deep learning approach for analysing dispersion relations in complex layered composite structures.
\end{abstract}
\end{frontmatter}
\section{Introduction}
Recent advances in machine learning, particularly deep learning, have significantly transformed computational modeling in science and engineering \cite{lecun2015deep,huang2022applications}. The development of efficient optimization algorithms and high-performance computational infrastructures has enabled neural networks to address increasingly complex physical problems with remarkable predictive capability \cite{goodfellow2016deep}. Among these developments, the integration of machine learning with governing physical laws has emerged as a promising paradigm for solving differential equation-based models \cite{karniadakis2021physics}.
Partial differential equations (PDEs) form the fundamental mathematical framework used to model a wide variety of physical phenomena, including wave propagation, fluid dynamics, heat transfer, and material deformation. In many realistic engineering systems, especially those involving multiphysics interactions, the governing behavior is represented by coupled systems of PDEs involving multiple interacting field variables. The numerical solution of such systems using traditional approaches, such as the finite element method (FEM) and finite difference method (FDM), often requires fine discretization, careful mesh generation, and substantial computational effort. These challenges become more pronounced in high-frequency regimes, strongly coupled systems, and heterogeneous layered media.
To overcome these challenges, Physics-Informed Neural Networks (PINNs) have been introduced as an effective computational approach. First introduced by Raissi et al. \cite{raissi2019physics}, PINNs embed the governing differential equations directly into the training process of neural networks through automatic differentiation. Instead of relying solely on data, the neural network is constrained by residuals of governing equations together with associated initial and boundary conditions.
The integration of physics-based modeling with machine learning techniques establishes a unified computational framework capable of solving complex boundary value problems without the need for explicit mesh generation.

Within the framework of PINNs, derivatives of the neural network outputs with respect to the input variables, such as spatial and temporal coordinates, are computed accurately using automatic differentiation \cite{raissi2019physics,jiang2024physics}. This capability enables the governing PDEs to be embedded into the training process of a neural network. 
The neural network parameters are optimized by minimizing a composite loss function that accounts for the governing equation residuals as well as the imposed boundary and initial conditions.
Through this physics-constrained optimization, the neural network learns to approximate solutions that satisfy both the available data and the underlying physical laws \cite{karniadakis2021physics}.
In practical implementations, data-driven PINNs where observational or simulated data are incorporated into the training procedure are more commonly employed than purely physics-based PINNs that rely solely on the governing equations. This preference arises because purely physics-constrained formulations may encounter challenges related to convergence speed and solution accuracy, particularly when dealing with complex or highly nonlinear systems \cite{haghighat2022physics}.

Since the introduction of the PINN framework, significant research efforts have been directed toward improving its robustness, efficiency, and predictive accuracy for solving complex scientific and engineering problems. Various studies have proposed strategies to enhance the training process and address optimization difficulties associated with PINNs. For instance, dynamic weighting strategies have been developed to balance the contributions of different components of the loss function during training \cite{xiang2022self}. In addition, adaptive activation functions have been introduced to improve the expressiveness of neural networks and facilitate better convergence behavior \cite{jagtap2022deep}. Sequential training approaches have also been explored to improve the stability and accuracy of PINN solutions, particularly for inverse and complex PDE problems \cite{amini2023inverse}.
Alongside improvements in training methodologies, several architectural variants of PINNs have been proposed to address specific computational challenges and expand the applicability of the framework. Notable examples include Bayesian Physics-Informed Neural Networks (BPINNs) \cite{yang2021b}, which incorporate uncertainty quantification into the learning process; Conservative PINNs (cPINNs) \cite{jagtap2020conservative}, which enforce conservation laws across domain interfaces; Extended PINNs (XPINNs) \cite{hu2021extended}, which enable domain decomposition for improved scalability; Parareal Physics-Informed Neural Networks (PPINNs) \cite{meng2020ppinn}, which integrate parallel-in-time algorithms with PINN training; Physics-Informed Generative Adversarial Networks (PI-GANs) \cite{yang2020physics}, which combine adversarial learning with physical constraints; and Gradient-Enhanced PINNs (gPINNs) \cite{eshkofti2023gradient}, which incorporate gradient information of the governing equations to improve solution accuracy. These developments aim to overcome the challenges of the standard PINN framework and enhance its capability to solve complex PDE-based problems across various scientific domains.

Despite the rapid development of PINNs and their successful application to a wide range of engineering and scientific problems, several complex physical phenomena governed by coupled partial differential equations remain challenging to analyze. One important area of study arises in the investigation of wave propagation in heterogeneous solid media. In particular, the propagation of shear waves in layered and anisotropic structures has received considerable attention due to its relevance in geophysics, earthquake engineering, and the analysis of advanced composite materials \cite{achenbach2012wave}.
In solid mechanics and materials science, orthotropic materials have attracted significant interest because of their important role in various engineering applications \cite{lekhnitskii1964theory}. These materials possess distinct mechanical properties along three mutually perpendicular directions and therefore provide a realistic representation of many natural and engineered structures. Several commonly used materials, such as wood and graphite–epoxy composites, exhibit orthotropic characteristics and are widely employed in fields such as aerospace engineering and structural engineering \cite{jones2018mechanics}. Moreover, wave propagation in orthotropic media has attracted considerable attention due to its applications in several fields such as optics, biology, acoustics, geophysics, seismic engineering, and electromagnetism.
Geophysical studies indicate that the Earth's interior, much like its external structure, consists of several distinct layers. This layered configuration has been inferred indirectly by examining the travel times of seismic waves that undergo reflection and refraction during earthquakes. As seismic waves propagate beneath the Earth's surface, they encounter different geological layers that often exhibit anisotropic behavior. Understanding these wave propagation phenomena is therefore essential for applications in exploration geophysics, geology, and seismology.
In many practical situations, materials are not perfectly homogeneous; instead, their mechanical properties vary gradually along certain spatial directions. Such materials are commonly referred to as functionally graded materials (FGMs), in which the gradual variation of material properties is introduced to enhance structural performance, reduce stress concentrations, and improve resistance to mechanical and thermal loads \cite{delfosse1998fundamentals}. When layered orthotropic structures with functional grading are subjected to external influences such as gravitational forces and pre-existing initial stresses, their dynamic response becomes considerably more complex. These factors can significantly influence the propagation characteristics of shear waves, including their velocity, attenuation, and dispersion behavior.

Because of the practical significance of layered anisotropic structures, wave propagation in multilayered media has been widely investigated over the past several decades. Early foundational studies focused on developing numerical and analytical techniques for analyzing wave motion in layered structures. Alterman and Karal \cite{alterman1968propagation} employed the finite difference method to investigate the propagation of elastic waves in layered media. Subsequently, Kelly et al. \cite{kelly1976synthetic} utilized a finite-difference approach to generate synthetic seismograms for studying seismic wave propagation. Theoretical investigations on wave propagation in multilayered orthotropic media were later presented by Nayfeh \cite{nayfeh1988theoretical}, providing important insights into the dynamic behavior of anisotropic layered structures.
Further developments in this field addressed more complex material configurations and wave phenomena. Hua \cite{hua2007love} studied Love wave propagation in layered graded composite structures with imperfectly bonded interfaces. Kalyani et al. \cite{kalyani2008finite} analyzed wave propagation in a monoclinic multilayered structure using Haskell’s matrix method and obtained phase and group velocities through a finite difference approach. Ahmed and Abo-Dahab \cite{ahmed2010propagation} investigated the propagation of Love waves in an orthotropic granular layer under initial stress overlying a semi-infinite granular medium. Chattopadhyay et al. \cite{chattopadhyay2010dispersion} examined the dispersion characteristics of shear waves in multilayered magnetoelastic self-reinforced media using numerical techniques.
More recent studies have continued to explore wave propagation in increasingly complex heterogeneous and composite structures. Gupta \cite{gupta2017implementation} investigated SH-wave propagation in a multilayered magnetoelastic orthotropic composite medium. Singh et al. \cite{singh2022propagation} examined the propagation characteristics of Love-type waves at an electro-mechanically imperfect interface in piezoelectric fiber-reinforced composite structures.
Chaudhary et al. \cite{chaudhary2025mechanics} investigated SH-wave propagation in viscoelastic structures using a mechanics-based framework that integrates machine learning techniques with analytical and numerical approaches.
Garg and Vaishnav \cite{garg2025love} studied Love-type wave characteristics in heterogeneous hydrogel layers over functionally graded piezoelectric fiber-reinforced composite substrates. Similarly, Gajroiya and Sikka \cite{gajroiya2025dynamics} analyzed the dynamics of Love-type wave propagation in composite transversely isotropic porous structures. These studies collectively highlight the continued interest in understanding wave propagation phenomena in layered and heterogeneous media.

Although numerous studies have investigated wave propagation in layered anisotropic media, most of the existing works have focused on specific material configurations or have relied on classical analytical and numerical approaches. For instance, Ahmed and Abo-Dahab \cite{ahmed2010propagation}, Nayfeh \cite{nayfeh1988theoretical}, and Gupta \cite{gupta2017implementation} analyzed wave propagation characteristics in various multilayered orthotropic media. Similarly, Kalyani et al. \cite{kalyani2008finite} and Chattopadhyay et al. \cite{chattopadhyay2010dispersion} examined shear wave propagation in layered structures using matrix-based and finite difference methods. More recent investigations have also considered wave propagation in complex heterogeneous and composite materials \cite{singh2022propagation, garg2025love, gajroiya2025dynamics}.
Despite these contributions, the combined influence of functional grading, orthotropic material behavior, gravitational effects, and initial stress on shear wave propagation in layered media remains relatively less explored. Furthermore, most of the existing studies rely on traditional analytical or numerical techniques, which may face limitations when dealing with highly heterogeneous and coupled systems. In this context, PINNs provide a promising alternative by integrating machine learning with governing physical laws to efficiently approximate solutions of complex PDE-based models.

To address the aforementioned challenges, the present study investigates the propagation of SH-waves in an initially stressed functionally graded magnetoelastic orthotropic layer resting on an initially stressed functionally graded orthotropic substrate under the influence of gravity. The governing mathematical model describing the physical system is first formulated analytically, and the corresponding dispersion relation governing the wave propagation characteristics is derived.
To further analyze the problem, a PINN framework is employed to approximate the solution of governing differential equations and compute corresponding dispersion characteristics. The neural network is trained using the Adam optimization algorithm to ensure efficient and stable convergence. In order to accurately represent the distinct material behaviors of  magnetoelastic orthotropic layer and functionally graded orthotropic half-space, different neural network architectures are employed for the two regions.
The accuracy and performance of the proposed computational framework are evaluated through a detailed error analysis. Various error metrics, including the $L_1$, $L_2$, and $L_\infty$ norms, together with relative and absolute errors, are computed. In addition, the influence of different activation functions on the predictive capability of the PINN model is examined. A parametric analysis is also conducted by varying the number of neurons and  hidden layers in the network architecture. The training procedure is carried out with a stopping criterion of $10^{4}$ epochs and an error tolerance of $10^{-6}$. The dispersion results obtained from the PINN framework are finally compared with the analytically derived dispersion relation in order to assess accuracy and effectiveness of proposed approach.

The organization of the manuscript is as follows. The analytical formulation of the problem is presented in Section \ref{Analytical Formulation}. Subsection \ref{Problem Formulation} describes the geometry of the physical model. Subsection \ref{Governing equation of motion and solution for the layer} develops the governing equation of motion and corresponding solution for the upper functionally graded magnetoelastic orthotropic layer, while Subsection \ref{Governing equation of motion and solution for the half-space} formulates the mathematical model for the underlying functionally graded orthotropic half-space. Subsection \ref{Boundary Conditions} describes the boundary conditions imposed on the system, and Subsection \ref{Dispersion relation} derives the dispersion relation for the considered problem.
Section \ref{Physics-Informed Neural Networks for Dispersion Analysis} introduces the formulation of the physics-informed neural network (PINN) framework employed for the dispersion analysis.  Section \ref{Numerical Simulation and Graphical Representation} provides the numerical simulations and graphical representations of the results. In this section, the dispersion relation obtained from the analytical formulation is analyzed and compared with the results obtained from the PINN model. The corresponding error analysis with respect to the dispersion relation is also presented for different neural network architectures. Furthermore, the influence of various material parameters on the dispersion characteristics is investigated. Section \ref{Conclusion} concludes the main findings of the study.
 \section{Analytical Formulation}
 \label{Analytical Formulation}
 \subsection{Problem formulation}
 \label{Problem Formulation}
\begin{figure}
     \centering
     \includegraphics[width=1\linewidth]{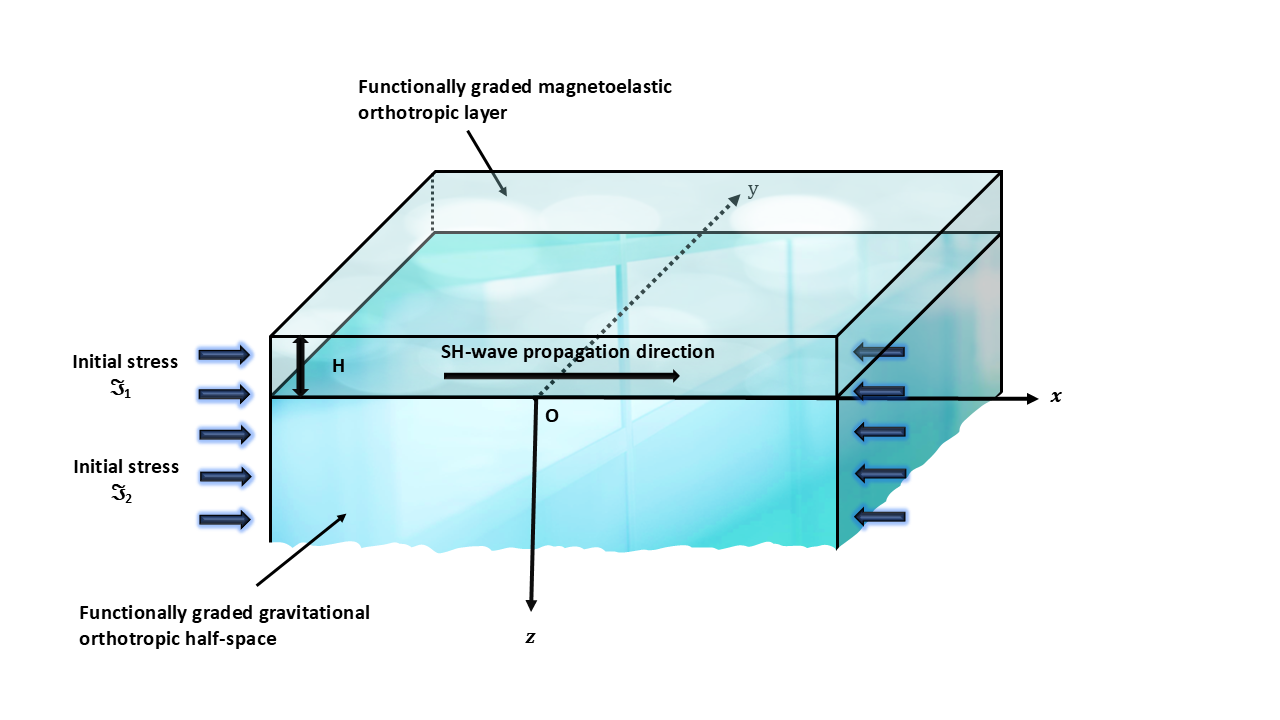}
    \caption{Schematic representation of the pre-stressed functionally graded layer–half-space composite medium}
     \label{geometery}
 \end{figure}

Consider the propagation of shear horizontal (SH) waves with phase velocity $c$ in a layered composite structure consisting of a pre-stressed functionally graded magnetoelastic orthotropic layer resting on a functionally graded orthotropic half-space under the influence of gravity. The cross-sectional configuration of the considered system is illustrated in Fig.~\ref{geometery}. 
The medium is subjected to horizontal initial stress, with $\Im_1$ acting along the $x$-direction in the upper layer and $\Im_2$ in the underlying half-space. The material properties in both regions vary continuously with depth due to functional grading, enabling mechanical parameters such as stiffness and damping to be tailored for applications including vibration mitigation and seismic wave attenuation.
To describe the physical configuration, a Cartesian coordinate system $O\text{-}xyz$ is introduced. The $x$-axis is aligned with the direction of wave propagation, whereas the $z$-axis represents the depth measured downward into the half-space. The parameter $H$ denotes the thickness of the orthotropic layer. Accordingly, the upper layer occupies the region $-H \leq z \leq 0$, while the lower medium extends infinitely in the positive $z$-direction and represents the half-space.
Let $(u_i, v_i, w_i)$ denote the displacement components of a material particle along the $x$-, $y$-, and $z$- directions, respectively. For SH-waves propagating in the $x$–$y$ plane with the direction of propagation along the $x$-axis, particle motion occurs only in the $y$-direction, while displacements in $x$- and $z$-directions remain negligible. Therefore, the displacement field corresponding to SH-wave propagation can be expressed as
\[
u_i = 0, \qquad w_i = 0, \qquad v_i = v_i(x,z,t), \qquad (i=1,2),
\]
where $i=1$ and $i=2$ refer to the upper layer and the underlying half-space, respectively.

\subsection{Governing equation of motion and solution for the layer}
\label{Governing equation of motion and solution for the layer}
The governing equation of motion for a functionally graded magnetoelastic orthotropic layer in the presence of initial stress are given by \cite{alam2018love,chaudhary2025mechanics}:
\be
\begin{aligned}
\frac{\partial \tau_{11}^{(1)}}{\partial x}
+ \frac{\partial \tau_{12}^{(1)}}{\partial y}
+ \frac{\partial \tau_{13}^{(1)}}{\partial z}
- \Im_1 \left(
\frac{\partial \omega_{y}}{\partial y}
- \frac{\partial \omega_{z}}{\partial z}
\right) + (\vec{J} \times \vec{B})_x
&= \rho_1 \frac{\partial^{2} u_1}{\partial t^{2}}, \\[8pt]
\frac{\partial \tau_{21}^{(1)}}{\partial x}
+ \frac{\partial \tau_{22}^{(1)}}{\partial y}
+ \frac{\partial \tau_{23}^{(1)}}{\partial z}
- \Im_1 \left(
\frac{\partial \omega_{z}}{\partial x}
\right) + (\vec{J} \times \vec{B})_y
&= \rho_1 \frac{\partial^{2} v_1}{\partial t^{2}}, \\[8pt]
\frac{\partial \tau_{31}^{(1)}}{\partial x}
+ \frac{\partial \tau_{32}^{(1)}}{\partial y}
+ \frac{\partial \tau_{33}^{(1)}}{\partial z}
- \Im_1 \left(
\frac{\partial \omega_{y}}{\partial x}
\right) + (\vec{J} \times \vec{B})_z
&= \rho_1 \frac{\partial^{2} w_1}{\partial t^{2}}.
\end{aligned}\label{eq1}
\ee 
Here, $\tau_{ij}^{(1)}$ denote stress components in the layer, $\rho_1$ is mass density, and $\Im_1$ is initial stress parameter, while $\omega_x$, $\omega_y$, and $\omega_z$ represent rotations about the $x$, $y$, and $z$ axes, respectively. The vector $\vec{J}$ denotes the electric current density, $\vec{B}$ represents the magnetic induction vector, and $(\vec{J} \times \vec{B})_i$ is the $i$th component of the Lorentz force $(\vec{J} \times \vec{B})$.\\
The constitutive relations for the orthotropic layer are given as  \cite{chaudhary2025mechanics}:
\be
\begin{aligned}
\tau^{(1)}_{11} &= \mu^{(1)}_{11} e_{11} + \mu^{(1)}_{12} e_{22} + \mu^{(1)}_{13} e_{33}, \\
\tau^{(1)}_{22} &= \mu^{(1)}_{21} e_{11} + \mu^{(1)}_{22} e_{22} + \mu^{(1)}_{23} e_{33}, \\
\tau^{(1)}_{33} &= \mu^{(1)}_{31} e_{11} + \mu^{(1)}_{32} e_{22} + \mu^{(1)}_{33} e_{33}, \\
\tau^{(1)}_{23} &= 2 \mu^{(1)}_{44} e_{23}, \\
\tau^{(1)}_{31} &= 2 \mu^{(1)}_{55} e_{31}, \\
\tau^{(1)}_{12} &= 2 \mu^{(1)}_{66} e_{12},
\end{aligned} \label{eq2} 
\ee
where $\mu^{(1)}_{ij}$ are the elastic parameters in upper layer.\\
The strain displacement relation is given as \cite{alam2018love}:
\be
\begin{aligned}
e_{11} &= \frac{\partial u_1}{\partial x}, \qquad
e_{12}  = \frac{\partial v_1}{\partial x}
+ \frac{\partial u_1}{\partial y},
\qquad
e_{22} = \frac{\partial v_1}{\partial y},\\
e_{23} &= \frac{\partial w_1}{\partial y}
+ \frac{\partial v_1}{\partial z},\qquad
e_{33}  = \frac{\partial w_1}{\partial z},\qquad
e_{13} = \frac{\partial w_1}{\partial x}
+ \frac{\partial u_1}{\partial z}.
\end{aligned}
\ee\label{eq3}
The rotational components along the $x$, $y$ and $z$ direction are given by \cite{majhi2025effect}
\be
\begin{aligned}
\omega_x &= \frac{1}{2}
\left(
\frac{\partial w_1}{\partial y}
-
\frac{\partial v_1}{\partial z}
\right), \qquad
\omega_y = \frac{1}{2}
\left(
\frac{\partial u_1}{\partial z}
-
\frac{\partial w_1}{\partial x}
\right), \qquad
\omega_z = \frac{1}{2}
\left(
\frac{\partial v_1}{\partial x}
-
\frac{\partial u_1}{\partial y}
\right).
\end{aligned}
\label{eq4}
\ee
To evaluate the Lorentz force contribution $(\vec{J} \times \vec{B})$ within the layer, the Maxwell equations governing the electromagnetic field are employed, as given in \cite{alam2018love}
\be
\nabla \cdot \vec{B} = 0, \qquad
\nabla \times \vec{E} = - \frac{\partial \vec{B}}{\partial t}, \qquad
\nabla \times \vec{M} = \vec{J}, \qquad
\vec{B} = \mu_e \vec{M},
\label{eq5}
\ee
where, the generalized Ohm's law is expressed as
\be
\vec{J} = \sigma \left( \vec{E} + \frac{\partial \vec{v}}{\partial t} \times \vec{B} \right)
\label{eq6}
\ee
and $\mu_e$ is the induced permeability; $\vec{M}$ includes both primary and induced magnetic field; $\sigma$ is the conduction coefficient; $\vec{E}$ is the induced electric field; $\vec{v} = (u_1,v_1,w_1)$ is the displacement vector for the layer.\\
Combining Eqs. (\ref{eq5}) and (\ref{eq6}) gives
\be
\frac{\partial \vec{M}}{\partial t}
=
\left[\frac{\nabla^2 \vec{M}}{\sigma \mu_e}
-
\nabla \times
\left(
\frac{\partial \vec{v}}{\partial t}
\times \vec{M}
\right)
\right].
\label{eq7}
\ee
Let $(M_1,M_2,M_3)$ and $(m_1,m_2,m_3)$; $m_i$ are the change in magnetic field components. 
Equation \ref{eq7} can be written in component form as follows
\be
\begin{aligned}
\frac{\partial M_1}{\partial t}
&= \frac{1}{\sigma \mu_e} \nabla^2 M_1, \\[6pt]
\frac{\partial M_2}{\partial t}
&= \frac{1}{\sigma \mu_e} \nabla^2 M_2
+ \frac{\partial}{\partial x}
\left(
M_1 \frac{\partial v_1}{\partial t}
\right)
+ \frac{\partial}{\partial z}
\left(
M_3 \frac{\partial v_1}{\partial t}
\right), \\[6pt]
\frac{\partial M_3}{\partial t}
&= \frac{1}{\sigma \mu_e} \nabla^2 M_3.
\label{eq8}
\end{aligned}
\ee
In the limiting case $\sigma \to \infty$, corresponding to a perfectly conducting medium, Eq. (\ref{eq8}) reduces to the following form
\be
\frac{\partial M_1}{\partial t}
=
0
=
\frac{\partial M_3}{\partial t}
\label{eq9},
\ee
and
\be
\frac{\partial M_2}{\partial t}=
\left[
\frac{\partial}{\partial x}
\left(
M_1 \frac{\partial v_1}{\partial t}
\right)
+
\frac{\partial}{\partial z}
\left(
M_3 \frac{\partial v_1}{\partial t}
\right)
\right].
\label{eq10}
\ee
Thus, Eq. (\ref{eq9}) shows that there is no perturbation in $M_1$ and $M_3$, whereas Eq. (\ref{eq10}) indicates the possibility of perturbation in $M_2$. Therefore, taking this small perturbation $m_2$ in $M_2$.
\begin{equation}
\begin{aligned}
M_1 &= M_{01}, \qquad
M_2 &= M_{02} + m_2, \qquad
M_3 &= M_{03},
\end{aligned}
\label{eq11}
\end{equation}
where $(M_{01}, M_{02}, M_{03})$ represent the components of the initial magnetic field $\vec{M}_0$, and initially $m_2 = 0$.
The magnetic field vector $\vec{M}$ can be written in component form as $\vec{M} = (M_0 \cos\phi,\, 0,\, M_0 \sin\phi)$, where $M_0 = |\vec{M}_0|$ denotes the magnitude of the magnetic field and $\phi$ represents the angle between the magnetic field and the direction of wave propagation.
Thus, 
\begin{equation}
\vec{M} = (M_0 \cos\phi,\, m_2,\, M_0 \sin\phi).
\label{eq12}
\end{equation}
Using Eq. (\ref{eq12}) in Eq. (\ref{eq10}), the following expression is obtained,
\begin{equation}
\frac{\partial m_2}{\partial t}
=
\frac{\partial}{\partial t}
\left[
M_0 \cos\phi \frac{\partial v_1}{\partial x}
+
M_0 \sin\phi \frac{\partial v_1}{\partial z}
\right].
\label{eq13}
\end{equation}
After integrating Eq. (\ref{eq13}) with respect to $t$, the following expression is obtained
\begin{equation}
m_2
=
M_0
\left[
\cos\phi \frac{\partial v_1}{\partial x}
+
\sin\phi \frac{\partial v_1}{\partial z}
\right].
\label{eq14}
\end{equation}
The following expression is obtained using Eqs. (\ref{eq5}) and (\ref{eq6})
\begin{equation}
\vec{J} \times \vec{B}
=
\mu_e
\left[
(\vec{M} \cdot \nabla)\vec{M}
-
\frac{1}{2}\nabla M^2
\right].
\label{eq15}
\end{equation}
The only non-zero component of Eq. (\ref{eq15}) can be expressed as 
\be
(\vec{J} \times \vec{B})_y = \mu_e M_0^2 \left( \cos^2\phi \frac{\partial^2 v_1}{\partial x^2} + \sin 2 \phi \frac{\partial^2 v_1}{\partial x \partial z} + \sin^2\phi \frac{\partial^2 v_1}{\partial z^2} \right). 
\label{eq16}
\ee 
Combining Eqs. (\ref{eq1}) and (\ref{eq16}) yields
\be
\frac{\partial \tau_{12}^{(1)}}{\partial x}
+ \frac{\partial \tau_{23}^{(1)}}{\partial z}
- \frac{\Im_1}{2}
\frac{\partial^2 v_1}{\partial x^2} +\mu_e M_0^2 \left( \cos^2\phi \frac{\partial^2 v_1}{\partial x^2} + \sin 2 \phi \frac{\partial^2 v_1}{\partial x \partial z} + \sin^2\phi \frac{\partial^2 v_1}{\partial z^2} \right)
= \rho_1 \frac{\partial^2 v_1}{\partial t^2} .
\label{eq17}
\ee 
Substituting Eq. (\ref{eq2}) in (\ref{eq17}), the following expression is obtained
\begin{equation}
    \frac{\partial}{\partial x}\left( \mu_{66}^{(1)}\frac{\partial v_1}{\partial x} \right)+\frac{\partial}{\partial z}\left( \mu_{44}^{(1)}\frac{\partial v_1}{\partial z} \right)-\frac{\Im_1}{2} \frac{\partial^2 v_1}{\partial x^2}+\mu_e M_0^2 \left( \cos^2\phi \frac{\partial^2 v_1}{\partial x^2} + \sin 2 \phi \frac{\partial^2 v_1}{\partial x \partial z} + \sin^2\phi \frac{\partial^2 v_1}{\partial z^2} \right)
= \rho_1 \frac{\partial^2 v_1}{\partial t^2} 
\label{stress strain 1}
\end{equation}
For the inhomogeneous functionally graded magnetoelastic layer, the material properties are assumed to vary exponentially with depth and are expressed as
\begin{equation}
\mu_{66}^{(1)}=\mu_{66}^{(1)^\prime}e^{\beta_1 z}, \quad
\mu_{44}^{(1)}=\mu_{44}^{(1)^\prime}e^{\beta_1 z}, \quad
\Im_1=\Im_1^{\prime}e^{\beta_1 z}, \quad
\rho_1=\rho_1^{\prime}e^{\beta_1 z}, \quad
\mu_e=\mu_e^{\prime}e^{\beta_1 z},
\label{eq18}
\end{equation}
where $\beta_1$ denotes the heterogeneity parameter of the functionally graded layer. The primed quantities $\mu_{66}^{(1)^\prime}$, $\mu_{44}^{(1)^\prime}$, $\Im_1^{\prime}$, $\rho_1^{\prime}$, and $\mu_e^{\prime}$ represent the corresponding reference material constants evaluated at $z=0$.
After the substitution of Eq. (\ref{eq18}) in Eq. (\ref{stress strain 1}), the following expression is obtained
\be
\begin{aligned}
  \mu_{66}^{(1)'} \frac{\partial^2 v_1}{\partial x^2} + \mu_{44}^{(1)'} \frac{\partial^2 v_1}{\partial z^2} + \mu_{44}^{(1)'} \beta_1 \frac{\partial v_1}{\partial z} - \frac{\Im_1'}{2} \frac{\partial^2 v_1}{\partial x^2}+\mu_e' M_0^2 \Big( \cos^2 \phi \frac{\partial^2 v_1}{\partial x^2} + \sin 2 \phi \frac{\partial^2 v_1}{\partial x \partial z}+\sin^2 \phi \frac{\partial^2 v_1}{\partial z^2} \Big) 
  = \rho_1' \frac{\partial^2 v_1}{\partial t^2}.
\end{aligned}  \label{eq19}
\ee
The time-dependent wave solution is assumed of the form
\begin{equation}
v_{1}(x,z,t)=V_{1}(z)e^{i(\omega t-kx)},
\label{eq20}
\end{equation}
where $V_1(z)$ is an unknown amplitude function, $\omega=kc$ denotes the angular frequency, $c$ is the phase velocity of the wave, and $k$ denote the wave number.
Substituting the solution given in Eq. (\ref{eq20}) into Eq. (\ref{eq19}) yields
\be
\frac{d^2 V_1}{d z^2} + c_1 \frac{d V_1}{dz} + c_2 V_1 = 0 ,\label{eq21}
\ee
where, 
\be
c_1 = \frac{\mu_{44}^{(1)^\prime} \beta_1 - \mu_e^\prime M_0^2 i k \sin 2 \phi}{\mu_{44}^{(1)^\prime} + \mu_e^\prime M_0^2 \sin^2 \phi} \qquad \text{and} \qquad
c_2 = \frac{k^2(-\mu_{66}^{(1)^\prime}+ \frac{\Im_1}{2} - \mu_e^\prime M_0^2 \cos^2 \phi + \rho^\prime_1 c^2)}{\mu_{44}^{(1)^\prime} + \mu_e^\prime M_0^2 \sin^2 \phi} .\nonumber
\ee
To eliminate the first derivative term in Eq.~(\ref{eq21}), the transformation 
$V_1(z)=e^{-\frac{c_1}{2}z}G_1(z)$ is introduced. Substitution of this expression into Eq.~(\ref{eq21}) removes the term $\frac{dV_1}{dz}$ and results in the following differential equation governing $G_1(z)$:
\be
\frac{d^2 G_1}{d z^2} + \gamma^2 G_1 = 0 ,
\label{eq22}
\ee
where, $\gamma^2 = c_2 - \frac{c_1^2}{4}$.
Solving Eq. (\ref{eq22}), the final expression for the magnetoelastic layer is given by
\be
v_1(z) = e^{-\frac{1}{2}c_1 z} \left( A_1 \cos \gamma z +A_2 \sin \gamma z \right) e^{i(\omega t - kx)}
\label{v1z}
\ee
where $A_1$ and $A_2$ represent arbitrary constants.
\subsection{Governing equation of motion and solution for the half-space}
\label{Governing equation of motion and solution for the half-space}
The equations of motion governing a functionally graded orthotropic half-space under the combined effects of gravity and initial stress are given by \cite{majhi2025effect,kumari2017propagation}:
\begin{equation}
\displaystyle
\frac{\partial \tau_{12}^{(2)}}{\partial x}
+\frac{\partial \tau_{23}^{(2)}}{\partial z}
-\frac{\Im_2}{2}\frac{\partial^2 v_2}{\partial x^2}
-\frac{\partial}{\partial x}\left(\frac{\rho_2 g z}{2}\frac{\partial v_2}{\partial x}\right)
-\frac{\partial}{\partial z}\left(\frac{\rho_2 g z}{2}\frac{\partial v_2}{\partial z}\right)
=\rho_2\frac{\partial^2 v_2}{\partial t^2},
\label{eq23}
\end{equation}
where $\tau_{ij}^{(2)}$ represent the stress tensor components, $\rho_2$ denotes the mass density, $\Im_2$ is the initial stress parameter, and $g$ represents the acceleration due to gravity.
The corresponding stress components are expressed as
\begin{equation}
\displaystyle
\begin{bmatrix}
\tau_{12}^{(2)} \\
\tau_{23}^{(2)}
\end{bmatrix}
=
\begin{bmatrix}
\mu_{66}^{(2)}\dfrac{\partial v_2}{\partial x} \\
\mu_{44}^{(2)}\dfrac{\partial v_2}{\partial z}
\end{bmatrix},
\label{eq24}
\end{equation}
where $\mu_{44}^{(2)}$ and $\mu_{66}^{(2)}$ denote the shear elastic constants associated with the orthotropic half-space.
An exponential variation of the material properties with depth is assumed for the functionally graded gravitational orthotropic half-space, which may be expressed as
\begin{equation}
\mu_{44}^{(2)}=\mu_{44}^{(2)^\prime}e^{\beta_2 z}, \quad
\mu_{66}^{(2)}=\mu_{66}^{(2)^\prime}e^{\beta_2 z}, \quad
\Im_2=\Im_2^{\prime}e^{\beta_2 z}, \quad
\rho_2=\rho_2^{\prime}e^{\beta_2 z},
\label{eq25}
\end{equation}
where $\beta_2$ denotes the heterogeneity parameter governing the functional variation of the material properties along the depth direction. The primed quantities $\mu_{44}^{(2)^\prime}$, $\mu_{66}^{(2)^\prime}$, $\Im_2^{\prime}$, and $\rho_2^{\prime}$ represent the corresponding reference material constants evaluated at $z=0$.
Substituting Eqs. (\ref{eq24}) and (\ref{eq25}) into Eq. (\ref{eq23}), the governing equation of motion for the orthotropic half-space reduces to
\begin{equation}
\begin{aligned}
\mu_{66}^{(2)^\prime} \frac{\partial^2 v_2}{\partial x^2} + \mu_{44}^{(2)^\prime} \frac{\partial^2 v_2}{\partial z^2} + \mu_{44}^{(2)^\prime} \beta_2 \frac{\partial v_2}{\partial z} - \frac{\Im_2^\prime}{2}  \frac{\partial^2 v_2}{\partial x^2} - \frac{\rho_2^\prime gz}{2} \frac{\partial^2 v_2}{\partial x^2}  - \frac{\rho_2^\prime g}{2} \Big( \beta_2 z \frac{\partial v_2}{\partial z} +\frac{\partial v_2}{\partial z} + z \frac{\partial^2 v_2}{\partial z^2} \Big) = \rho_2^\prime \frac{\partial^2 v_2}{\partial t^2} 
\label{eq26}
\end{aligned}
\end{equation}
Assuming a time-dependent wave solution for the half-space in the form
\begin{equation}
v_2(x,z,t)=V_2(z)e^{i(\omega t-kx)},
\label{eq27}
\end{equation}
where $V_2(z)$ is an unknown amplitude function. Substituting Eq. (\ref{eq27}) into Eq. (\ref{eq26}), the governing equation reduces to
\begin{equation}
\frac{d^2 V_2}{dz^2}
+
\frac{\left(\mu_{44}^{(2)^\prime}\beta_2-\frac{\rho_2^{\prime}gz}{2}\beta_2-\frac{\rho_2^{\prime}g}{2}\right)}
{\mu_{44}^{(2)^\prime}-\frac{\rho_2^{\prime}gz}{2}}
\frac{dV_2}{dz}
+
\frac{k^2\left(-\mu_{66}^{(2)^\prime}+\frac{\Im_2^{\prime}}{2}+\frac{\rho_2^{\prime}gz}{2}+\rho_2^{\prime}c^2\right)}
{\mu_{44}^{(2)^\prime}-\frac{\rho_2^{\prime}gz}{2}}V_2=0.
\label{eq28}
\end{equation}
By applying the transformation
\begin{equation}
V_2(z)=\frac{G_2(z)}{\sqrt{\mu_{44}^{(2)^\prime}-\frac{\rho_2^{\prime}gz}{2}}}
\,e^{-\frac{\beta_2 z}{2}},
\end{equation}
Eq. (\ref{eq28}) may be expressed in the following form:
\begin{equation}
\frac{d^2 G_2(z)}{dz^2}
+
\left[
\frac{k^2\left(-\mu_{66}^{(2)^\prime}+\frac{\Im_2^{\prime}}{2}+\frac{\rho_2^{\prime}gz}{2}+\rho_2^{\prime}c^2\right)+\frac{\beta_2\rho_2^{\prime}g}{4}}
{\mu_{44}^{(2)^\prime}-\frac{\rho_2^{\prime}gz}{2}}
+
\frac{1}{16}\frac{\rho_2^{\prime 2}g^2}
{\left(\mu_{44}^{(2)^\prime}-\frac{\rho_2^{\prime}gz}{2}\right)^2}
-\frac{\beta_2^2}{4}
\right]G_2(z) = 0,
\label{eq29}
\end{equation}
By introducing the parameters $\sigma_1$ and $s$, Eq. (\ref{eq29}) reduces to the Whittaker differential equation \cite{spencer1974boundary} of the form
\begin{equation}
\frac{d^2 G_2(z)}{d\sigma_1^2}
+
\left[
-\frac{1}{4}
+\frac{s}{\sigma_1}
+\frac{1}{4\sigma_1^2}
\right] G_2(z) = 0,
\label{eq30}
\end{equation}
where
\begin{equation}
\sigma_1 =
\frac{2\beta_2}{\rho_2^{\prime} g}
\sqrt{1+\frac{4k^2}{\beta_2^2}}
\left(
\mu_{44}^{(2)^\prime}-\frac{\rho_2^{\prime} g z}{2}
\right),
\end{equation}
\begin{equation}
s=
\frac{2}{\rho_2^{\prime} g \beta_2
\sqrt{1+\frac{4k^2}{\beta_2^2}}}
\left[
k^2
\left(
-\mu_{66}^{(2)^\prime}
+\mu_{44}^{(2)^\prime}
+\frac{\Im_2^{\prime}}{2}
+\rho_2^{\prime} c^2
\right)
+\frac{\beta_2 \rho_2^{\prime} g}{4}
\right].
\end{equation}
Thus, the general solution corresponding to Eq. (\ref{eq30}) may be written as
\begin{equation}
G_2(\sigma_1)
=
A_3 W_{-s,0}(-\sigma_1)
+
A_4 W_{s,0}(\sigma_1),
\label{eq31}
\end{equation}
where $A_3$ and $A_4$ are arbitrary constants, and $W_{-s,0}(-\sigma_1)$ and $W_{s,0}(\sigma_1)$ denote the Whittaker functions.
The approximate solution of Eq. (\ref{eq31}), obtained by applying the radiation condition 
$G_2(\sigma_1) \rightarrow 0$ as $z \rightarrow \infty$, reduces to
\begin{equation}
G_2(\sigma_1)=A_3 W_{-s,0}(-\sigma_1).
\label{eq32}
\end{equation}
Thus, wave propagation in the orthotropic gravitational half-space involves only a single non-zero displacement component. The displacement field can therefore be expressed as
\begin{equation}
v_2(z) = A_3
\left(\mu_{44}^{(2)^\prime}-\frac{\rho_2^{\prime}gz}{2}\right)^{-1/2}
W_{-s,0}\!\left[
\sqrt{1+\frac{\beta_2^2}{4k^2}}
\left(\frac{4}{G}-2kz\right)
\right]
e^{-\frac{\beta_2 z}{2}},
\label{eq33}
\end{equation}
where 
\[
G=\frac{g\rho_2^{\prime}}{k\mu_{44}^{(2)^\prime}}
\]
denotes Biot's gravity parameter.

For physically admissible wave propagation in the half-space, the radiation condition $\mathrm{Im}(R_3)>0$ must be satisfied, where
\[
R_3^2=\frac{c^2\rho_2^{\prime}}{\mu_{44}^{(2)^\prime}-\rho_2^{\prime}}.
\]


\subsection{Boundary conditions}
\label{Boundary Conditions}
In the study of wave propagation, the displacement and stress fields must satisfy suitable boundary conditions at the surface boundary and at the material interface to maintain physical consistency. These conditions ensure that the system remains dynamically stable and free from discontinuities.
\begin{itemize}
    \item \textbf{Free surface of functionally graded magnetoelastic orthotropic layer} $(z=-H)$ \\
    Since the boundary condition is in contact with air or vacuum, it experiences no applied tangential traction; hence the shear stress component must vanish, leading to a stress free boundary condition.
    \begin{equation}
        \tau_{23}^{(1)}=0.
        \label{eq35}
    \end{equation}
    \item \textbf{Interface between layer and half-space} $(z=0)$\\
    To maintain the physical continuity, both the displacement and shear stress must remain continuous at the interface. This prevents displacement jumps and unbalanced forces:
    \begin{equation}
        v_1=v_2, \qquad \tau_{23}^{(1)}=\tau_{23}^{(2)}.
        \label{eq36}
    \end{equation}
\end{itemize}

\subsection{Dispersion relation}
\label{Dispersion relation}
The dispersion relation is obtained by applying the boundary conditions given in Eqs. (\ref{eq35}) and (\ref{eq36}) to Eqs. (\ref{eq2}), (\ref{v1z}), (\ref{eq24}), and (\ref{eq33}). This leads to the following three equations:
\be
e^{\left(\frac{1}{2} c_1 H \right)} \left[ \left(- \frac{1}{2} c_1 \cos \gamma H - \gamma \sin \gamma H \right)A_1 + \left( \gamma \cos \gamma H - \frac{1}{2} c_1 \sin \gamma H \right)A_2 \right] = 0,
\label{eq37}
\ee
\be
\left. A_1 - \frac{A_3}{\sqrt{\mu_{44}^{(2)^\prime}}} W_{-s,0}\left[\sqrt{1 + \frac{\beta_2^2}{4K^2}}
\left(\frac{4}{G}-2kz \right)\right] \right|_{z=0} =0,
\label{eq38}
\ee
\be
\begin{aligned}
 \frac{1}{2} c_1 \mu_{44}^{(1)\prime} A_1 
 - \mu_{44}^{(1)^\prime} \gamma A_2 
 + A_3 \mu_{44}^{(2)^\prime} 
 \left\{ 
 W_{-s,0}\left[
 \sqrt{1 + \frac{\beta_2^2}{4K^2}}
\left. \left(\frac{4}{G}-2kz \right)
 \right] \right|_{z=0}
 \left( 
 \frac{\rho_2^\prime g - 2 \mu_{44}^{(2)^\prime} \beta_2}
 {4 \left(\mu_{44}^{(2)^\prime}\right)^{3/2}}  
 \right) \right. \\
 \qquad \left.
 + \frac{1}{\sqrt{\mu_{44}^{(2)^\prime}}} 
 \frac{d}{dz}  
 W_{-s,0}\left[
 \sqrt{1 + \frac{\beta_2^2}{4K^2}}
\left. \left(\frac{4}{G}-2kz \right)
 \right]\right|_{z=0}
 \right\} = 0.
 \label{eq39}
\end{aligned}
\ee
The arbitrary constants $A_1$, $A_2$, and $A_3$ are eliminated from Eqs. (\ref{eq37})--(\ref{eq39}) to obtain

\begin{equation}
    \frac{1}{2}c_1\mu_{44}^{(1)^\prime}-\mu_{44}^{(1)^\prime}\gamma\left\{\frac{\frac{1}{2}c_1\cos\gamma H+\gamma\sin\gamma H}{\gamma\cos\gamma H-\frac{1}{2}c_1\sin\gamma H} \right\}+\left(\frac{\rho_2^\prime g-2 \mu_{44}^{(2)^\prime}\beta_2}{4} \right)+\mu_{44}^{(2)^\prime} X=0
    \label{eq40},
\end{equation}
where, 
\begin{equation}
\begin{aligned}
X =
\frac{
\left.
\dfrac{d}{dz}
W_{-s,0}\!\left[
\sqrt{1 + \frac{\beta_2^2}{4K^2}}
\left(\frac{4}{G} - 2kz \right)
\right]
\right|_{z=0}
}{
\left.
W_{-s,0}\!\left[
\sqrt{1 + \frac{\beta_2^2}{4K^2}}
\left(\frac{4}{G} - 2kz \right)
\right]
\right|_{z=0}
}.
\end{aligned}
\end{equation}
The asymptotic expansion \cite{spencer1974boundary} of Whittaker's function $\left.
W_{-s,0}\!\left[
\sqrt{1 + \frac{\beta_2^2}{4K^2}}
\left(\frac{4}{G} - 2kz \right)
\right]
\right|_{z=0}$ and \\ $\left.
\dfrac{d}{dz}
W_{-s,0}\!\left[
\sqrt{1 + \frac{\beta_2^2}{4K^2}}
\left(\frac{4}{G} - 2kz \right)
\right]
\right|_{z=0}$ for large argument (retaining upto second term) is given by
\begin{equation}
\begin{aligned}
    \left. W_{-s,0} \left(\sqrt{1+\frac{\beta_2^2}{4k^2}}\left(\frac{4}{G}-2kz \right) \right) \right|_{z=0} \sim e^{\left(\sqrt{1+\frac{\beta_2^2}{4k^2}} \frac{2}{G}  \right)} \left(\sqrt{1+\frac{\beta_2^2}{4k^2}} \frac{4}{G} \right)^{-s} \left\{1- \frac{(s+0.5)^2}{\sqrt{1+\frac{\beta_2^2}{4k^2}} \frac{4}{G}} \right\},
    \label{eq42}
\end{aligned}
\end{equation}
\begin{equation}
\begin{aligned}
\left.\frac{dW_{-s,0}}{dz}\!\left(\sqrt{1+\frac{\beta_2^2}{4k^2}}
\left(\frac{4}{G}-2kz\right)\right)\right|_{z=0}
&\sim
e^{\!\left(\sqrt{1+\frac{\beta_2^2}{4k^2}}\frac{2}{G}\right)}
\left(\sqrt{1+\frac{\beta_2^2}{4k^2}}\frac{4}{G}\right)^{-s}
\left(-k\sqrt{1+\frac{\beta_2^2}{4k^2}}\right) \\[6pt]
&\quad \times
\left\{\left(1-\frac{(s+0.5)^2}
{\sqrt{1+\frac{\beta_2^2}{4k^2}}\frac{4}{G}}\right)
\left(1-\frac{2s}
{\sqrt{1+\frac{\beta_2^2}{4k^2}}\frac{4}{G}} \right)
+\frac{2(s+0.5)^2}
{\left(1+\frac{\beta_2^2}{4k^2}\right)\left(\frac{4}{G}\right)^2}
\right\}.
\label{eq43}
\end{aligned}
\end{equation}
By means of Eqs. (\ref{eq40})--(\ref{eq43}), it follows that
\begin{equation}
    \frac{1}{2}c_1\mu_{44}^{(1)^\prime}-\mu_{44}^{(1)^\prime}\gamma \left\{\frac{\frac{1}{2}c_1\cos\gamma H+\gamma\sin\gamma H}{\gamma\cos\gamma H-\frac{1}{2}c_1\sin\gamma H} \right\}+\left(\frac{\rho_2^\prime g-2 \mu_{44}^{(2)^\prime}\beta_2}{4} \right)+\mu_{44}^{(2)^\prime} Q=0
    \label{eq44},
\end{equation}
where, 
\begin{equation}
    Q=-k\left\{\sqrt{1+\frac{\beta_2}{4k^2}}-2s+\frac{(s+0.5)^2G}{2\left(\sqrt{1+\frac{\beta_2^2}{4k^2}}\frac{4}{G}-(s+0.5)^2\right)} \right\}.
\end{equation}
The equation \ref{eq44} gives the required dispersion equation of SH-wave propagating in magneto-elastic orthotropic layer lying above a functionally graded gravitational orthotropic half-space.


\section{Physics-Informed Neural Networks for Dispersion Analysis}
\label{Physics-Informed Neural Networks for Dispersion Analysis}
This section presents the formulation of a physics-informed neural network (PINN) framework for computing the dispersion characteristics of SH-waves propagating in a functionally graded layered structure resting on a functionally graded half-space under the influence of gravity. The governing equations obtained in the previous section consist of coupled variable-coefficient partial differential equations that incorporate material heterogeneity, initial stress, and electromagnetic body forces. For generality, the system may be expressed in the operator form
\begin{equation}
\mathcal{F}\big(\mathbf{Y}(\mathbf{Z});\boldsymbol{\psi}\big)=0, 
\quad \mathbf{Z}\in\Omega,
\label{eq:pinn_pde}
\end{equation}
\begin{equation}
\mathcal{B}\big(\mathbf{Y}(\mathbf{Z})\big)=\mathbf{g}(\mathbf{Z}),
\quad \mathbf{Z}\in\partial\Omega,
\label{eq:pinn_bc}
\end{equation}
where $\mathcal{F}(\cdot)$ denotes the differential operator corresponding to the reduced governing equations of SH-wave propagation, and $\mathcal{B}(\cdot)$ represents the boundary and interface operators enforcing the stress-free surface condition, displacement continuity, traction continuity at the layer–half-space interface, and the far-field decay requirement. The vector $\mathbf{Y}$ contains the unknown field variables, namely the complex amplitude of the displacement function in both the layer and the half-space, while $\boldsymbol{\psi}$ represents the deterministic model parameters including elastic constants, density, grading parameters, magnetic field intensity, gravitational acceleration, and wave number. The spatial domain $\Omega$ consists of the graded layer and the underlying half-space, whereas $\partial\Omega$ includes the top surface, the material interface, and the truncated far-field boundary.
Although an analytical formulation of the dispersion relation may be derived through classical wave propagation theory, the simultaneous presence of functional grading, magnetoelastic coupling, and gravity-induced heterogeneity renders closed-form solutions highly involved. The PINN framework is therefore introduced to approximate the solution field $\mathbf{Y}(\mathbf{Z})$ by neural network surrogate $\hat{\mathbf{Y}}(\mathbf{Z};\boldsymbol{\theta})$, where $\boldsymbol{\theta}$ denotes the set of trainable parameters comprising weights and biases. In addition to approximating the displacement field, the unknown phase velocity $c$ is treated as an auxiliary trainable parameter, thereby transforming the dispersion relation into an eigenvalue identification problem within the PINN optimization framework.

\subsection{Neural network approximation}
\label{Neural Network Approximation}
A fully connected feed-forward neural network is adopted to approximate the depth-dependent displacement amplitude in both functionally graded layer and the underlying half-space. Under the time-harmonic assumption $v(x,z,t)=V(z)e^{i(\omega t-kx)}$, the unknown field is expressed in terms of the complex amplitude function $V(z)$. To facilitate numerical implementation within a real-valued computational framework, the complex amplitude is decomposed into its real and imaginary components as
\begin{equation}
V(z) = V_R(z) + i V_I(z),
\end{equation}
where $V_R(z)$ and $V_I(z)$ denote the real and imaginary parts, respectively.
Accordingly, the spatial coordinate $z$ is used as the network input, while the outputs correspond to the pair $\{V_R(z), V_I(z)\}$. This real-imaginary decomposition enables an efficient and stable enforcement of the coupled differential equations arising from the complex eigenvalue formulation of the dispersion problem. For the half-space, a similar representation is employed to ensure continuity at the interface separating the two materials.
The required first- and second-order spatial derivatives of the network output with respect to the input coordinate are evaluated using automatic differentiation (AD) provided by modern machine learning libraries.
Through backpropagation and computational graph construction, AD enables exact evaluation of derivatives up to machine precision. Consequently, the strong-form differential operators appearing in the governing equations and boundary/interface conditions can be directly enforced within the loss functional without introducing discretization error associated with classical numerical differentiation schemes. This preserves the continuous structure of the underlying wave propagation problem while leveraging the expressive capability of neural network approximators.

\subsection{Physics loss function}
\label{Physics Loss Function}
Since no labeled input--output data are available, the PINN training process relies exclusively on enforcing the governing physics. Accordingly, the loss function is formulated to include the residuals of the governing equations, boundary conditions, interface continuity conditions, far-field decay requirement, and eigenvalue normalization constraint. The total loss function is formulated as
\begin{equation}
\mathcal{L}_{\text{total}}(\boldsymbol{\theta},c)=
w_{\text{PDE}}\mathcal{L}_{\text{PDE}}+
w_{\text{TOP}}\mathcal{L}_{\text{TOP}}+
w_{\text{IC}}\mathcal{L}_{\text{IC}}+
w_{\text{FF}}\mathcal{L}_{\text{FF}}+
w_{\text{AMP}}\mathcal{L}_{\text{AMP}},
\label{eq:loss_total}
\end{equation}
where $c$ denotes the unknown phase velocity treated as a trainable parameter, and $w_{\text{PDE}}$, $w_{\text{TOP}}$, $w_{\text{IC}}$, $w_{\text{FF}}$, and $w_{\text{AMP}}$ are weighting coefficients introduced to balance the relative contributions of the individual components.

\noindent{\textit{Governing Equation Residual:}}
The residuals of the governing equations in both the functionally graded layer and the underlying half-space are incorporated into the PDE loss term as
\begin{equation}
\mathcal{L}_{\text{PDE}}=
\frac{1}{N_r}
\sum_{i=1}^{N_r}
\left\|
\mathcal{F}\big(\hat{\mathbf{Y}}(\mathbf{Z}_i;\boldsymbol{\theta},c)\big)
\right\|^2,
\label{eq:loss_pde}
\end{equation}
where $N_r$ corresponds to the number of collocation points employed in the computational domain.

\noindent{\textit{Boundary Condition at Top:}}
The traction-free condition at the top surface of the layer is enforced through
\begin{equation}
\mathcal{L}_{\text{TOP}}=
\frac{1}{N_{top}}
\sum_{i=1}^{N_{top}}
\left\|
\mathcal{B}_{\text{top}}\big(\hat{\mathbf{Y}}(\mathbf{Z}_i;\boldsymbol{\theta})\big)
\right\|^2,
\label{eq:loss_bc}
\end{equation}
where $N_{top}$ denotes the number of collocation points selected at the free surface.

\noindent{\textit{Interface Continuity:}}
Continuity of displacement and traction across the layer--half-space interface is enforced by
\begin{equation}
\mathcal{L}_{\text{IC}}=
\frac{1}{N_{ic}}
\sum_{i=1}^{N_{ic}}
\left\|
\mathcal{B}_{\text{int}}\big(\hat{\mathbf{Y}}(\mathbf{Z}_i;\boldsymbol{\theta})\big)
\right\|^2,
\label{eq:loss_ic}
\end{equation}
where $N_{ic}$ represents the number of collocation points located at the interface.

\noindent{\textit{Far-Field Decay Condition:}}
To ensure physically admissible wave attenuation in the half-space, a decay condition is imposed through
\begin{equation}
\mathcal{L}_{\text{FF}}=
\frac{1}{N_{ff}}
\sum_{i=1}^{N_{ff}}
\left\|
\hat{\mathbf{Y}}(\mathbf{Z}_i;\boldsymbol{\theta})
\right\|^2,
\label{eq:loss_ff}
\end{equation}
where $N_{ff}$ denotes the number of collocation points selected sufficiently far from the interface.

\noindent{\textit{Amplitude Normalization:}}
Since the governing equations constitute a homogeneous eigenvalue problem, the trivial solution must be excluded. To prevent collapse of the solution toward zero, a normalization condition is imposed at a selected reference location (e.g., the free surface),
\begin{equation}
\mathcal{L}_{\text{AMP}}=
\left|V_R(z_s)-1\right|^2
+
\left|V_I(z_s)\right|^2,
\label{eq:loss_amp}
\end{equation}
where $z_s$ denotes the top surface coordinate. This constraint fixes the amplitude and phase of the eigenfunction without altering the physical dispersion characteristics.

\subsection{Extraction of the dispersion relation}
\label{Extraction of the Dispersion Relation}
For a prescribed wave number $k$, the phase velocity $c=\omega/k$ is treated as an additional trainable parameter within the PINN framework. The dispersion relation is therefore obtained by reformulating the wave propagation problem as an eigenvalue identification task. During training, both neural network parameters $\boldsymbol{\theta}$ and phase velocity $c$ are optimized simultaneously by minimizing the total loss function defined in Eq. \eqref{eq:loss_total}.

The physically admissible dispersion relation corresponds to those values of $c$ for which the governing equations, boundary conditions, interface continuity requirements, far-field decay condition, and amplitude normalization constraint are satisfied simultaneously. In other words, dispersion curve is identified as the set of phase velocities for which the strong-form residual of the boundary value problem approaches zero within the admissible SH-wave velocity window.

To construct the complete dispersion curve, the above optimization procedure is repeated for a sequence of prescribed wave numbers $\{k_j\}_{j=1}^{N_k}$. For each $k_j$, the trained phase velocity $c_j$ is recorded, yielding a discrete representation of the dispersion relation
\begin{equation}
c = c(k).
\end{equation}
This sweeping strategy allows the PINN framework to recover the dispersion characteristics without explicitly deriving the classical transcendental frequency equation, thereby providing a flexible computational alternative for complex heterogeneous media.

\subsection{Neural network architecture}
\label{Neural Network Architecture}
Fully connected feed-forward neural networks are employed to approximate the displacement amplitudes appearing in the dispersion analysis. The spatial coordinate in the thickness direction, $z$, is used as the single network input. Two independent neural networks are constructed: one corresponding to the functionally graded layer and one corresponding to the underlying half-space.

\noindent -For the functionally graded layer, the displacement field is represented in complex form following the harmonic wave assumption. Accordingly, the network outputs consist of two neurons corresponding to the real and imaginary components of the displacement amplitude, denoted by $V_R(z)$ and $V_I(z)$. This representation enables stable enforcement of the coupled real–imaginary formulation of the eigenvalue problem.

\noindent -For the half-space, a real-valued formulation is adopted, and a single-output neural network is employed to represent the decaying displacement field. This formulation is sufficient due to the monotonic attenuation of SH-waves within the half-space.

\begin{figure}
    \centering
    \includegraphics[width=1\linewidth]{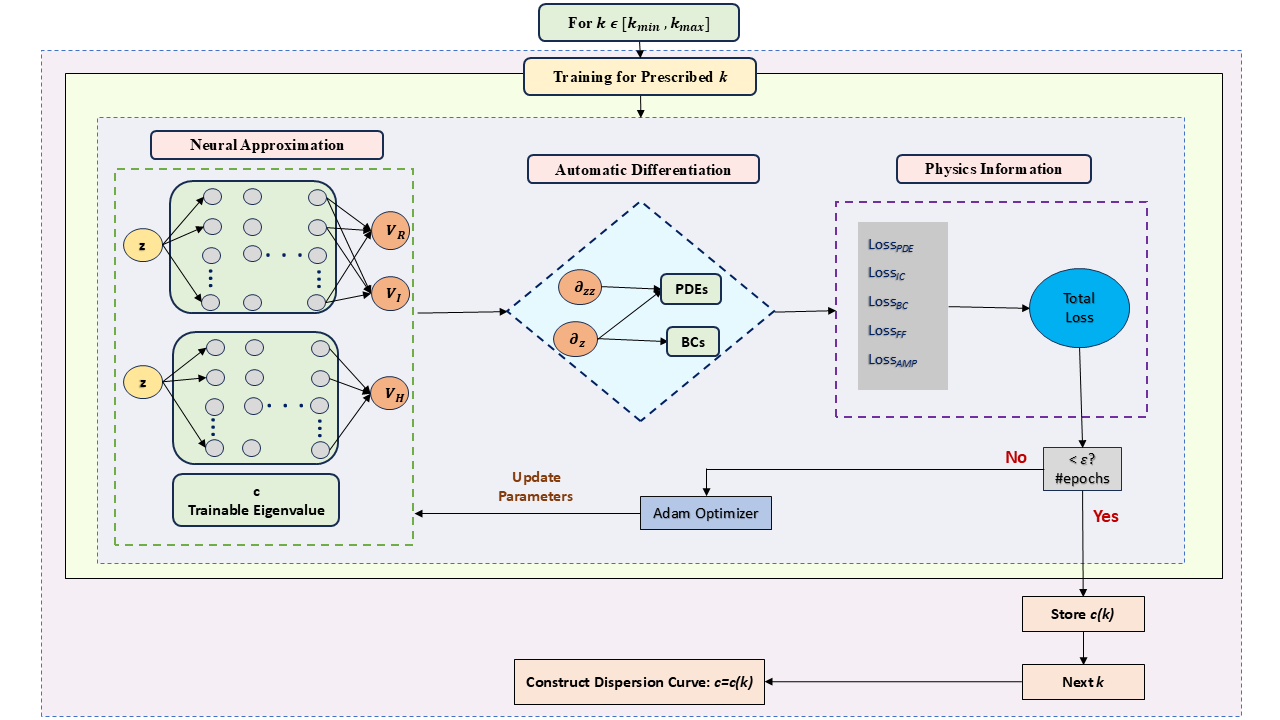}
    \caption{Schematic representation of the proposed PINN architecture for SH-wave dispersion analysis.}
    \label{fig:placeholder}
\end{figure}

 All hidden layers are fully connected and utilize the hyperbolic tangent ($\tanh$) activation function. The $\tanh$ activation function is chosen because of its smooth and infinitely differentiable nature, which facilitates the accurate evaluation of first- and second-order spatial derivatives through automatic differentiation. The output layer is kept linear to avoid imposing artificial restrictions on the solution amplitude.
Automatic differentiation, implemented within the computational framework, is used to evaluate all required spatial derivatives of the network outputs. 
 The detailed architectural parameters and computational settings used in the present PINN framework are summarized in Table~\ref{tab:architecture}. The overall structure of the proposed PINN model is illustrated schematically in Fig.~\ref{fig:placeholder}.

\begin{table}[h!]
\centering
\caption{Computational setup and neural network configuration used in the PINN-based dispersion analysis.}
\renewcommand{\arraystretch}{1.2}
\begin{tabular}{lcc}
\hline
\textbf{Parameter} & \textbf{Layer Network} & \textbf{Half-space Network} \\
\hline
Input variable & $z$ & $z$ \\
Input dimension & 1 & 1 \\
Output dimension & 2 ($V_R,V_I$) & 1 ($V$) \\
Neurons per layer & 50 & 50 \\
Hidden layers & 8 & 8 \\
Activation function & $\tanh$ & $\tanh$ \\
Output activation & Linear & Linear \\
Optimizer & \multicolumn{2}{c}{Adam} \\
Learning rate (networks) & \multicolumn{2}{c}{$10^{-4}$} \\
Trainable parameter & \multicolumn{2}{c}{Phase velocity $c$ (scalar)} \\
Epochs per wavenumber & \multicolumn{2}{c}{10000} \\
Stopping criterion & \multicolumn{2}{c}{Loss $<10^{-6}$ or 10000 epochs} \\
Training strategy & \multicolumn{2}{c}{Sequential training for each $k$ with continuation initialization} \\
\hline
\multicolumn{3}{c}{\textbf{Computational platform}} \\
\hline
Processor & \multicolumn{2}{c}{12th Gen Intel Core i5-1235U (1.30 GHz CPU)} \\
RAM & \multicolumn{2}{c}{16 GB} \\
Implementation & \multicolumn{2}{c}{Python with PyTorch} \\
\hline
\end{tabular}
\label{tab:architecture}
\end{table}

\section{Numerical Simulation and Graphical Representation}
\label{Numerical Simulation and Graphical Representation}
This section presents the numerical simulations carried out to analyze dispersion behaviour of shear waves and to evaluate the influence of different material parameters on the phase velocity. The analytical dispersion relation derived in the previous section is numerically evaluated using \textsc{MATLAB}. In parallel, the physics-informed neural network (PINN) framework is implemented in \textsc{Visual Studio} to compute the corresponding dispersion curves and validate the analytical predictions.
Moreover, a comparison between analytical solution and  PINN-based results is performed to evaluate the accuracy and robustness of the proposed computational method.
For the purpose of numerical computation, the material parameters adopted in the simulations are taken from the work of S.K.~Panja \textit{et al.}~\cite{panja2022propagation}. The corresponding elastic constants and related parameters employed in the present study are listed in Table~\ref{tab:elastic_constants}.
\begin{table}[h!]
\centering
\caption{Parameter}
\label{tab:elastic_constants}
\begin{tabular}{|c|c|c|}
\hline
\textbf{Parameters} & \textbf{Orthotropic layer} & \textbf{Orthotropic half-space} \\ \hline
$\rho$ (kg/m$^{3}$) & $9890 $ & $3400$\\ \hline
$\mu_{44}$ (Pa) & $4.35 \times 10^{9}$ & $5.3 \times 10^{9}$ \\ \hline
$\mu_{66}$ (Pa) & $5 \times 10^{9}$ & $6.47 \times 10^{9}$ \\ \hline
$g$(m/s$^2$)& $-$&$9.81$\\ \hline
\end{tabular}
\label{table 2}
\end{table}

\subsection{Dispersion characteristics}

\begin{figure}[htbp]
\centering

\begin{subfigure}[b]{0.47\textwidth}
\includegraphics[width=\textwidth]{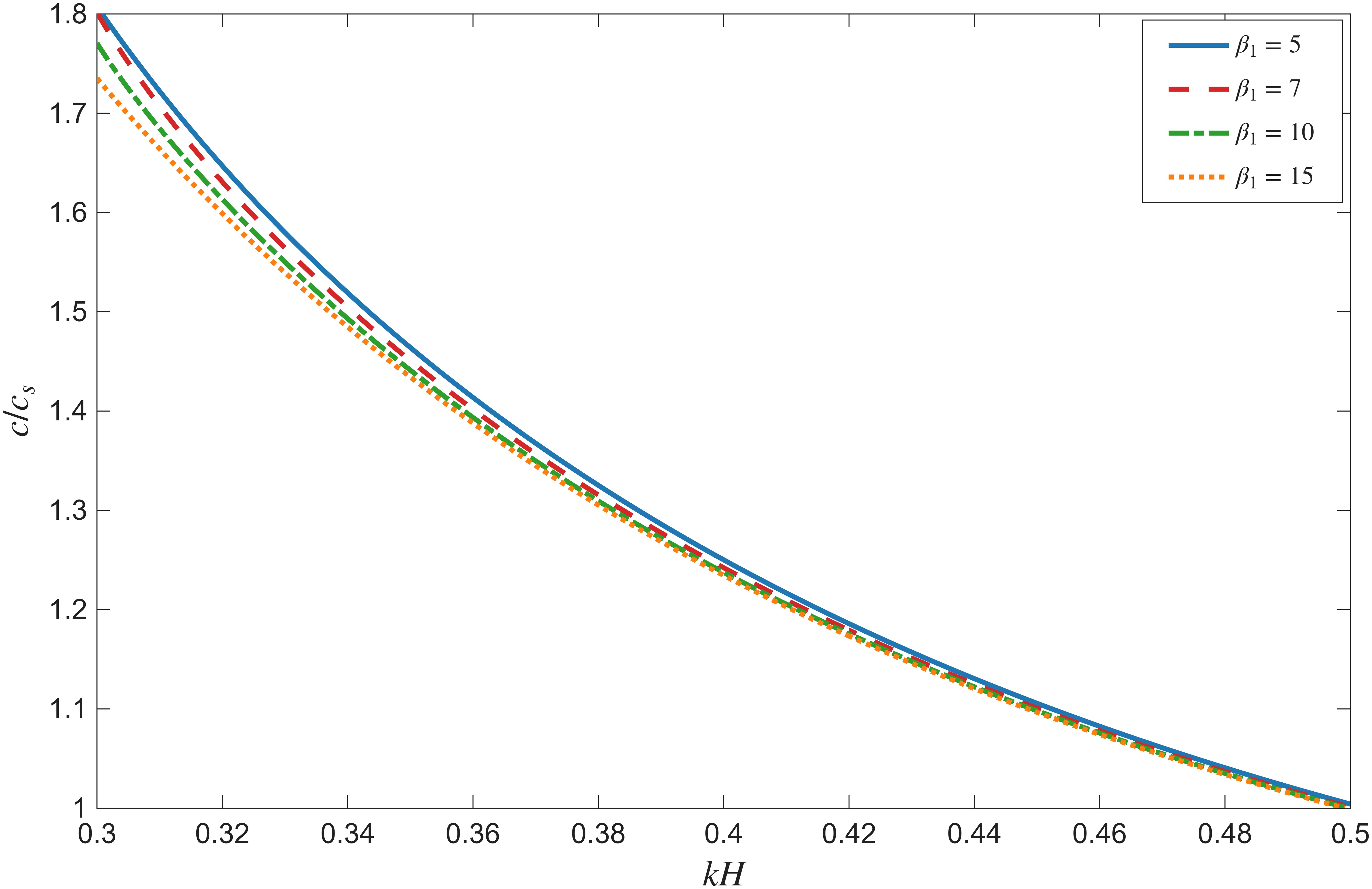}
\caption{}
\label{(2a)}
\end{subfigure}
\hfill
\begin{subfigure}[b]{0.47\textwidth}
\includegraphics[width=\textwidth]{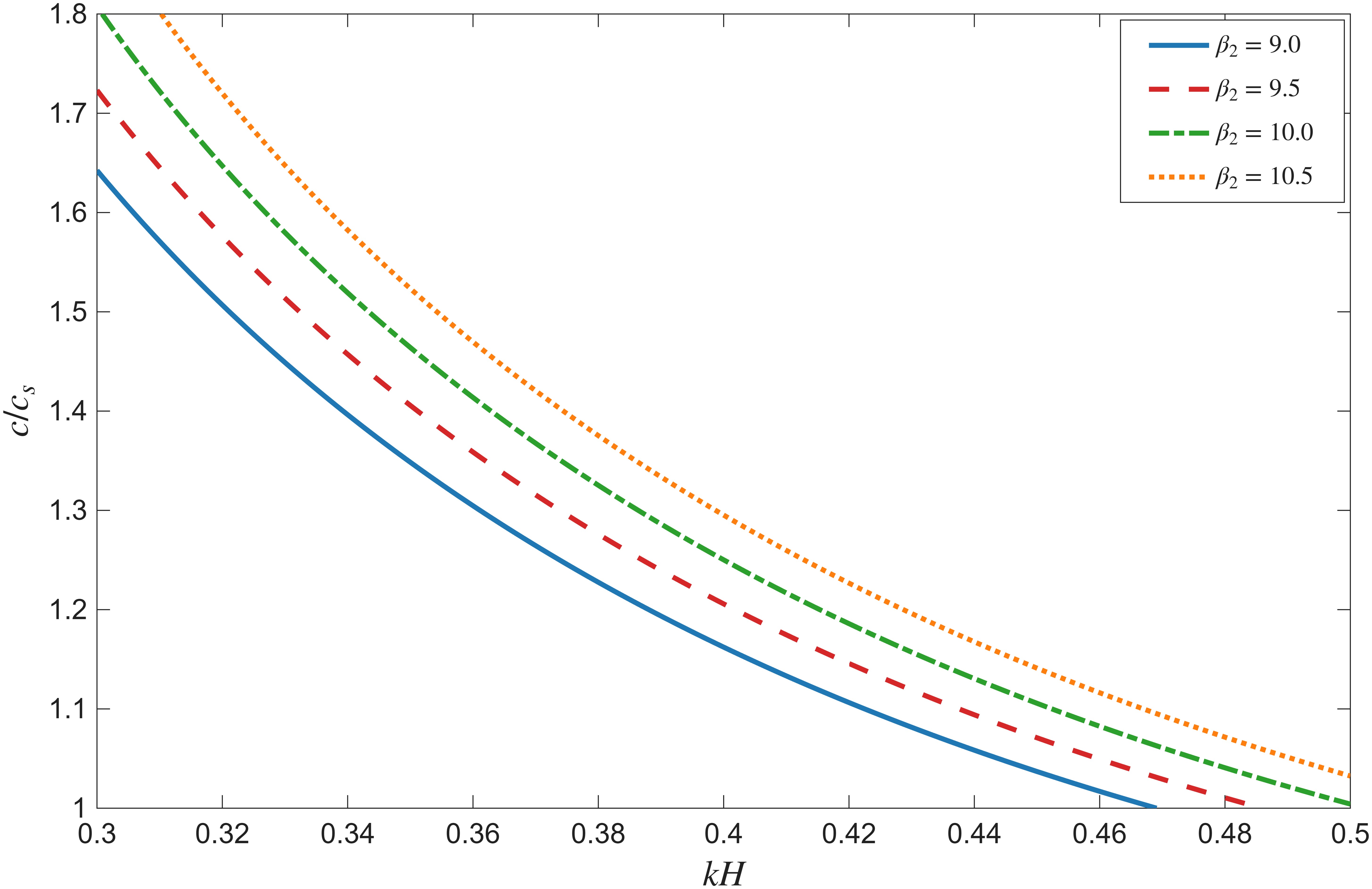}
\caption{}
\label{(2b)}
\end{subfigure}
\caption{Dependence of normalized phase velocity on the dimensionless wavenumber $kH$ for different values of the heterogeneity parameter $\beta_i$ (a) in the layer and (b) in the half-space.}
\label{Figure 2}
\end{figure}

\begin{figure}[htbp]
\centering

\begin{subfigure}[b]{0.47\textwidth}
\includegraphics[width=\textwidth]{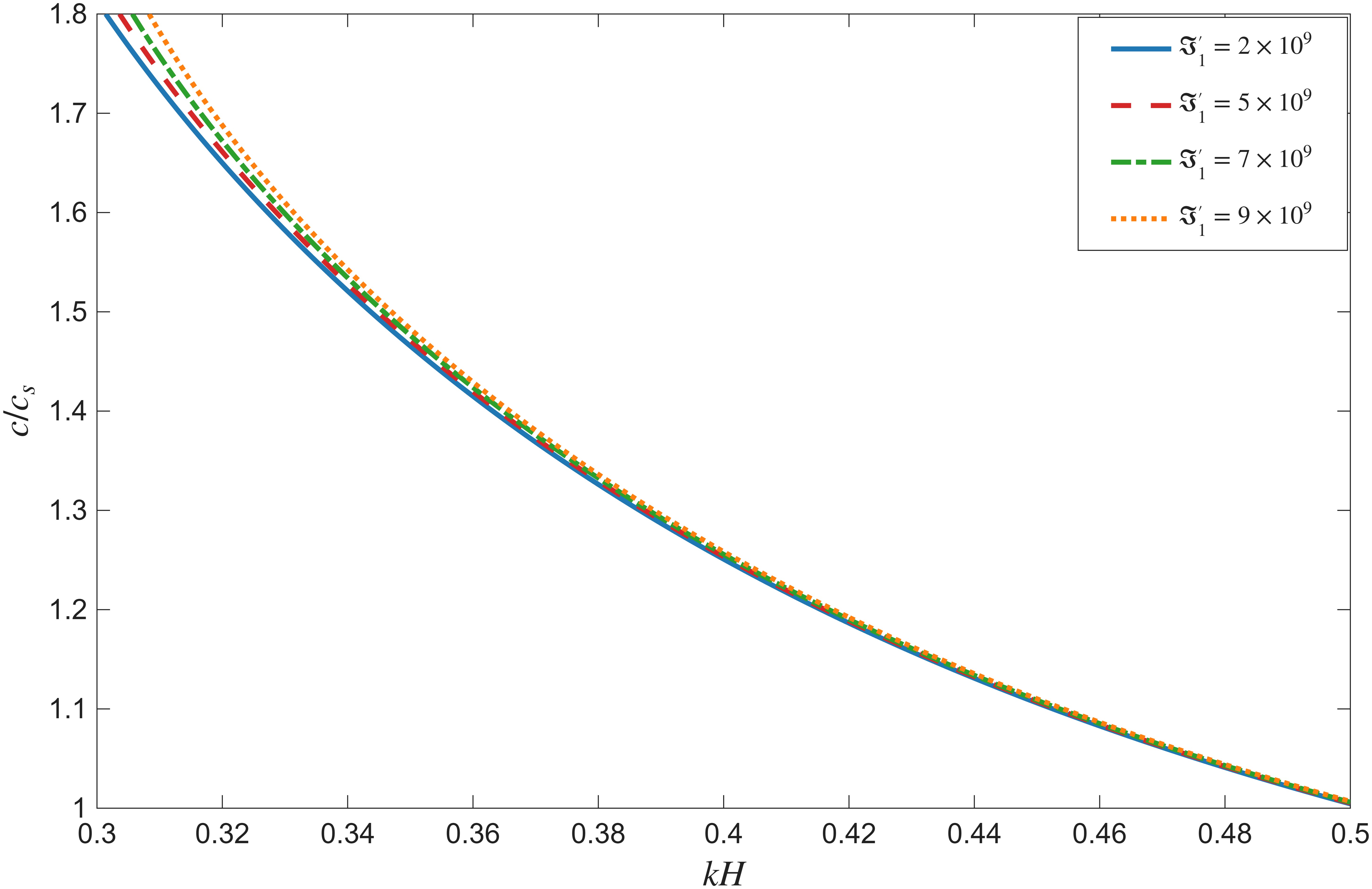}
\caption{}
\label{(3a)}
\end{subfigure}
\hfill
\begin{subfigure}[b]{0.47\textwidth}
\includegraphics[width=\textwidth]{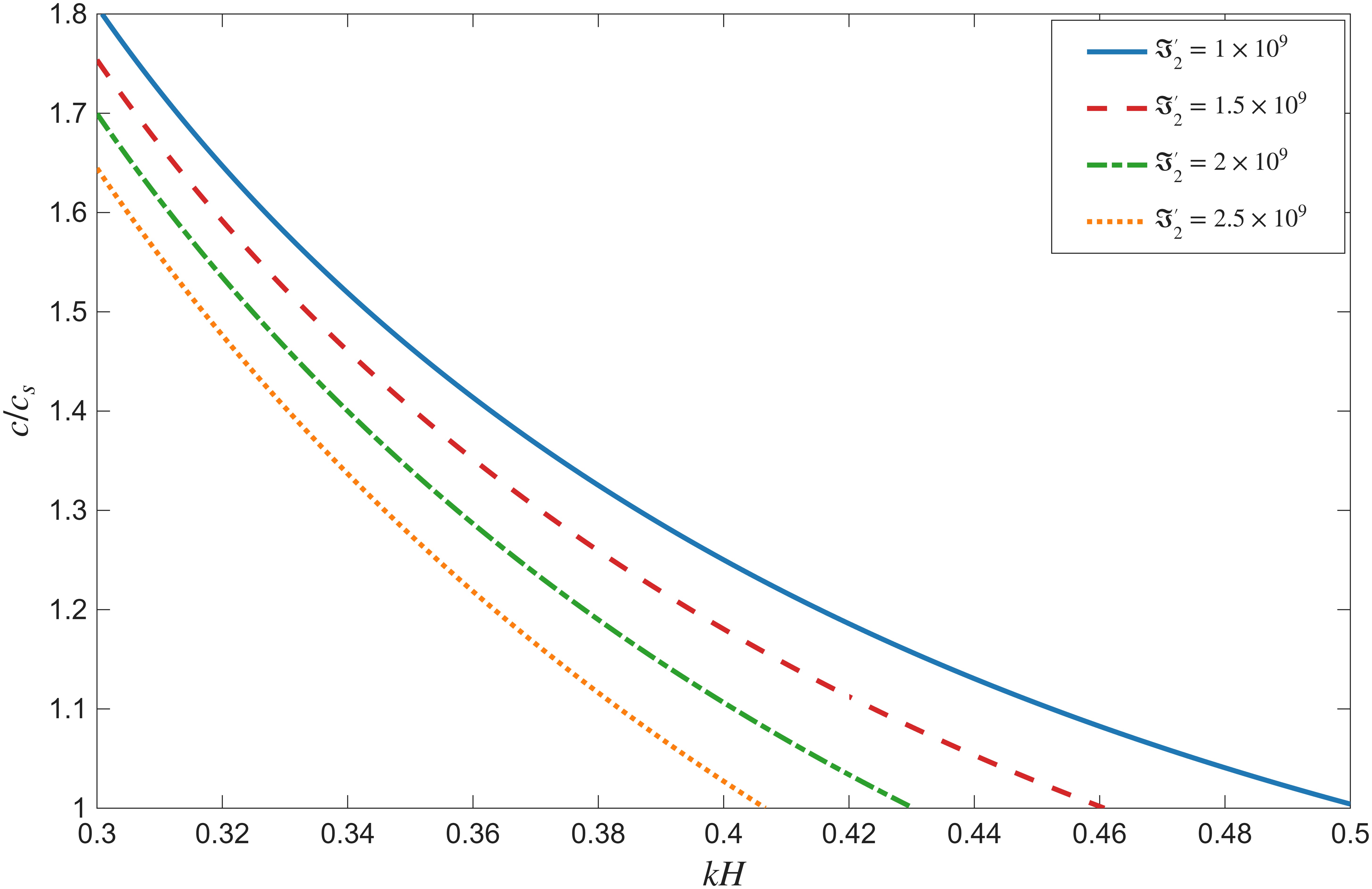}
\caption{}
\label{(3b)}
\end{subfigure}
\caption{Dependence of normalized phase velocity on the dimensionless wavenumber $kH$ for different initial stress values in (a) the layer and (b) the half-space.}
\label{Figure 3}
\end{figure}
\begin{figure}[htbp]
\centering
\begin{subfigure}[b]{0.47\textwidth}
\includegraphics[width=\textwidth]{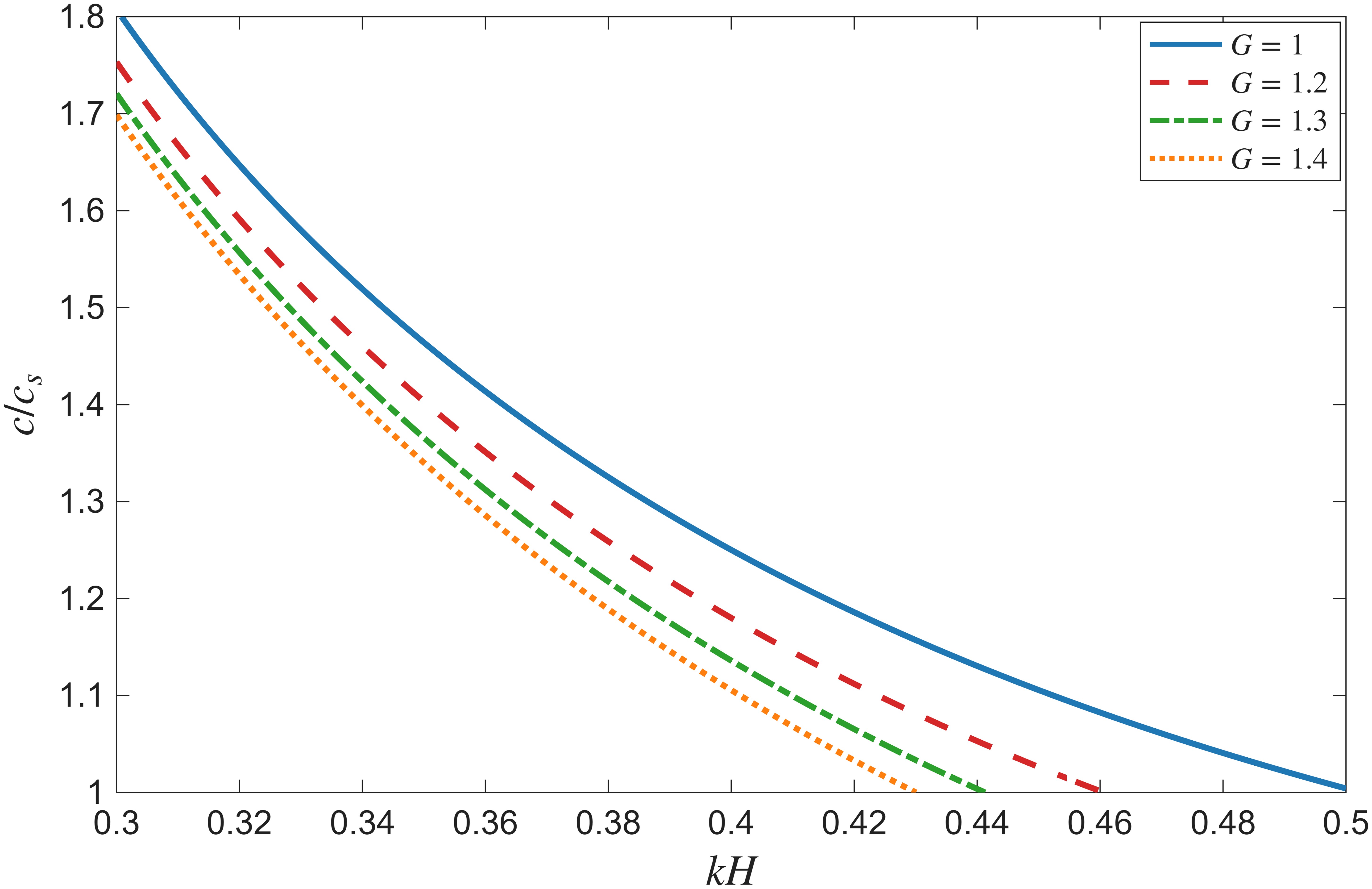}
\caption{}
\label{(4a)}
\end{subfigure}
\hfill
\begin{subfigure}[b]{0.47\textwidth}
\includegraphics[width=\textwidth]{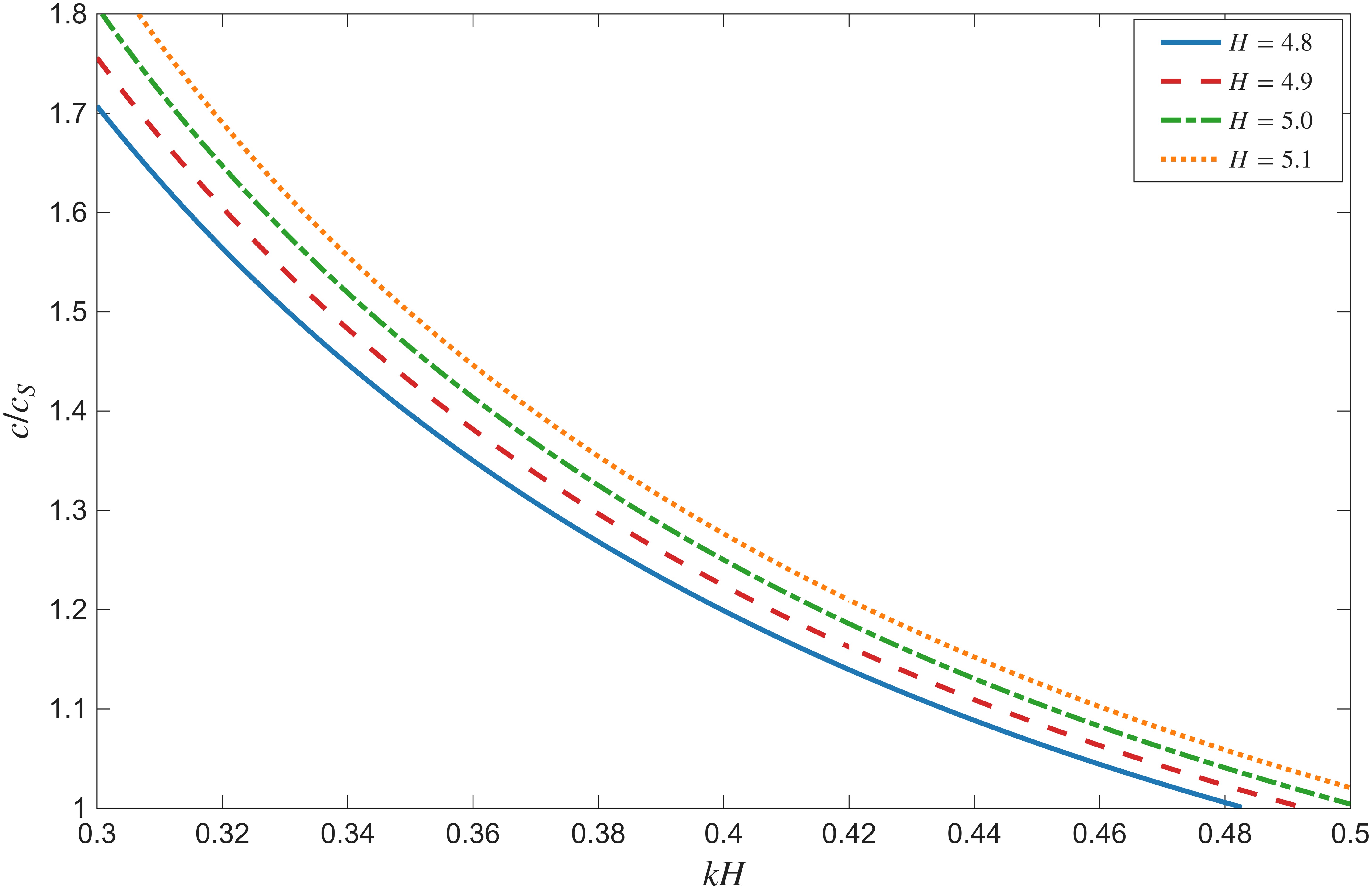}
\caption{}
\label{(4b)}
\end{subfigure}
\caption{Dependence of normalized phase velocity on the dimensionless wavenumber $kH$ for different values of (a) Biot’s gravity parameter $(G)$ and (b) layer height $(H)$.}
\label{Figure 4}
\end{figure}
\begin{figure}[htbp]
\centering
\begin{subfigure}[b]{0.47\textwidth}
\includegraphics[width=\textwidth]{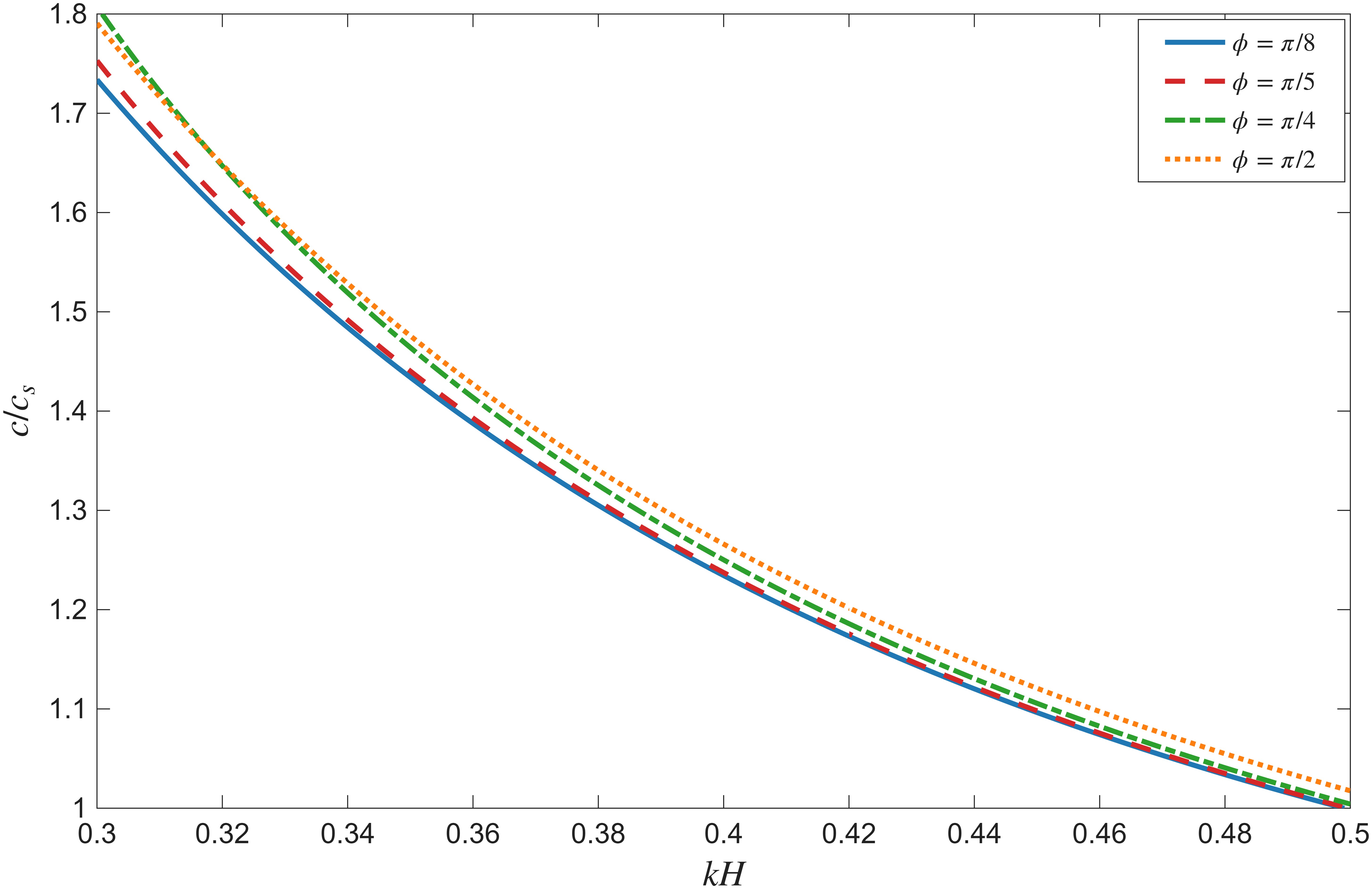}
\caption{}
\label{(5a)}
\end{subfigure}
\hfill
\begin{subfigure}[b]{0.47\textwidth}
\includegraphics[width=\textwidth]{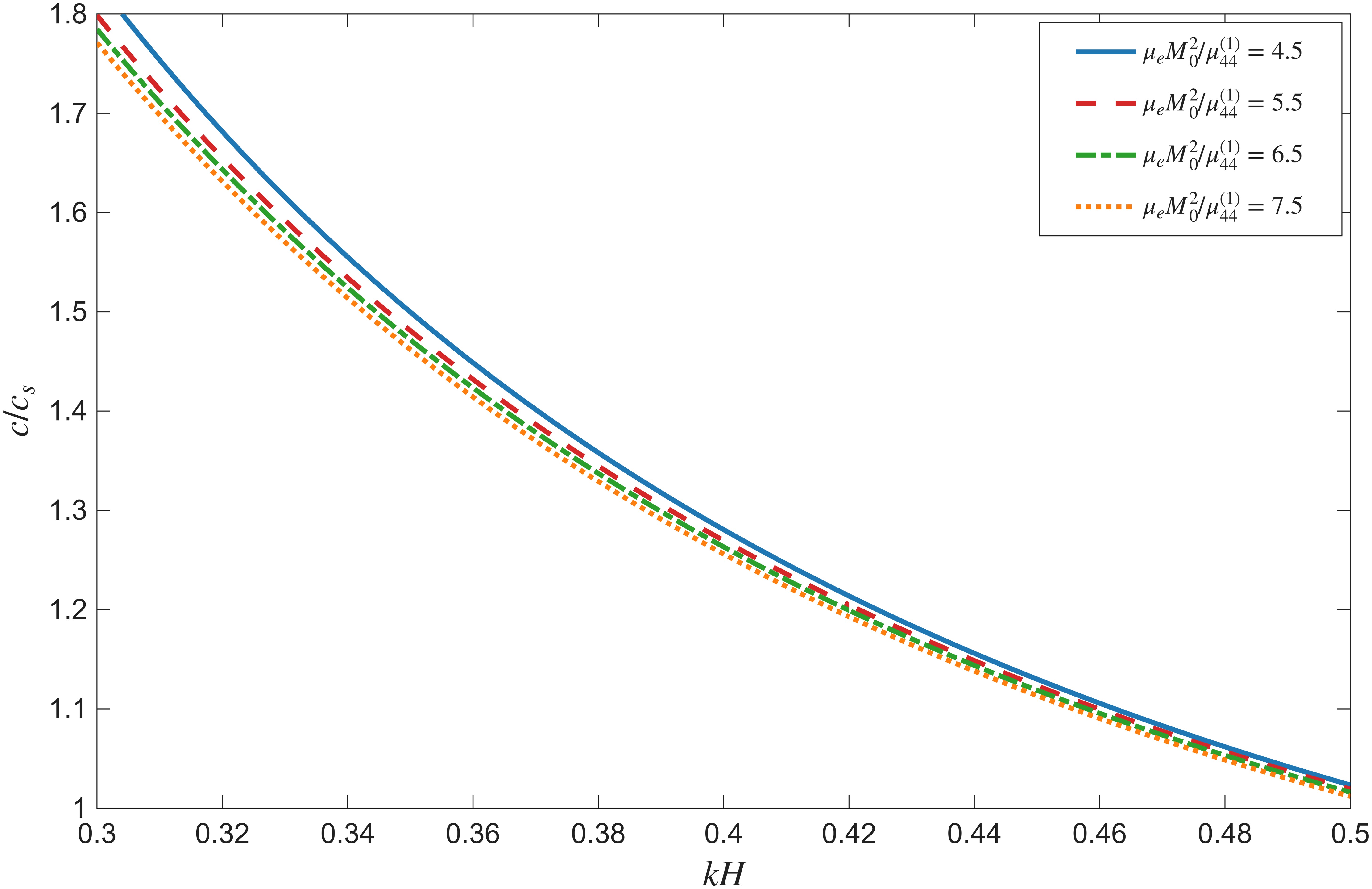}
\caption{}
\label{(5b)}
\end{subfigure}
\caption{Dependence of normalized phase velocity on the dimensionless wavenumber $kH$ for different values of (a) the magnetic field angle $(\phi)$ and (b) the magnetic permeability parameter $\mu_e M_0^2/\mu_{44}^{(1)}$.}
\label{Figure 5}
\end{figure}
Figures \ref{Figure 2}-\ref{Figure 5} illustrate the influence of various material parameters on the normalized phase velocity $c/c_s$ of SH-waves as a function of the dimensionless wave number $kH$, where $c_s=\sqrt{\mu_{44}^{(1)}/\rho_1}$. 
Figure \ref{Figure 2} depicts the variation of the normalized phase velocity with the dimensionless wave number $kH$ for different values of the heterogeneity parameter in the layer and the half-space. The results are obtained for the Biot gravity parameter $G=1$, layer thickness $H=5$, and magnetic field angle $\phi=\frac{\pi}{4}$.
Figure \ref{(2a)} shows that increasing heterogeneity in the layer reduces its effective stiffness, leading to a decrease in phase velocity. Conversely, Figure \ref{(2b)} shows that increasing heterogeneity in the half-space enhances its effective stiffness, resulting in an increase in phase velocity.
Figure \ref{Figure 3} illustrates the variation of the normalized phase velocity with the dimensionless wave number $kH$ for different values of the initial stress applied in (a) the layer $(\Im_1^\prime)$ and (b) the half-space $(\Im_2^\prime)$. The computations are carried out for heterogeneity parameters $\beta_1=5$ and $\beta_2=10$, magnetic field angle $\phi=\frac{\pi}{4}$, and Biot gravity parameter $G=1$.
From Fig.~\ref{(3a)}, it is observed that the phase velocity increases with increasing initial stress in the layer. This indicates that the presence of initial stress enhances the effective shear rigidity of the layer, leading to an increase in its stiffness.
Figure \ref{(3b)} shows that the phase velocity decreases as the initial stress in the half-space increases. This behavior indicates that the initial stress significantly influences the mechanical response of the substrate by altering its effective stiffness. Consequently, the supporting medium becomes relatively softer, which weakens the guided wave mode and leads to a reduction in the propagation speed.
Figure \ref{Figure 4} demonstrates the significant influence of Biot's gravity parameter ($G$) and the layer thickness ($H$) on the phase velocity of SH waves. Figure \ref{(4a)} shows that the phase velocity decreases as the value of Biot's gravity parameter increases.
From Fig.~\ref{(4b)}, it is observed that the phase velocity increases with increasing layer thickness. This behaviour can be attributed to improved confinement of wave energy within the layer, which reduces its leakage into the underlying half-space. Figure \ref{Figure 5} presents the variation of phase velocity with the wave number for different values of (a) the magnetic field angle $(\phi)$ and (b) the induced permeability.
From Fig.~\ref{(5a)}, it is observed that the phase velocity increases with the magnetic field angle $(\phi)$. This trend indicates that an increase in $\phi$ enhances the effective shear stiffness of the guiding layer, which strengthens the confinement of SH-wave energy within the layer.
Figure \ref{(5b)} illustrates a reduction in phase velocity with increasing induced permeability $\mu_e M_0^2/\mu_{44}^{(1)}$. This behavior suggests that the electromagnetic coupling modifies the mechanical response of the medium, reducing its effective rigidity. As a result, the supporting medium becomes relatively softer, weakening the guided wave mode and consequently lowering the propagation speed.

\subsection{Validation of the PINN model}
\begin{figure}
    \centering
    \includegraphics[width=0.7\linewidth]{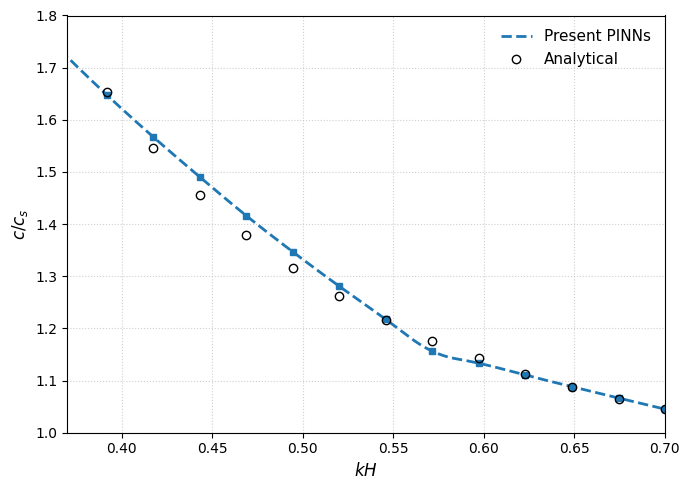}
\caption{Comparison of the normalized phase velocity $c/c_s$ obtained using the proposed PINN method and the analytical solution as a function of the dimensionless wavenumber $kH$.}
    \label{PINN vs Analytical}
\end{figure}

To assess the accuracy and reliability of the proposed PINN framework, the predicted dispersion characteristics are compared with analytical solution obtained in Section~\ref{Analytical Formulation}.
The comparison is presented in terms of the normalized phase velocity $c/c_s$ plotted against the dimensionless wave number $kH$.
For the numerical implementation of the PINN formulation, the material, electromagnetic, and geometric parameters used in simulations are summarized in Table~\ref{tab:pinn_params}.
\begin{table}[h!]
\centering
\caption{Material and geometric parameters used in the PINN-based dispersion analysis.}
\renewcommand{\arraystretch}{1.15}
\setlength{\tabcolsep}{4pt}
\begin{tabular}{cccccccccccc}
\hline
$\Im_1^\prime$ & $\beta_1$ & $\mu_e M_0^2/\mu_{44}^{(1)}$ & $\phi$ (rad) & $\Im_2^\prime$ & $\beta_2$ & $g$ (m/s$^2$) & $H(m)$ & $L$ & $k_{\min}$ & $ k_{\max}$ & $N_k$ \\
\hline
$1.0\times10^{9}$ & $5$ & $0.3$ & $\pi/6$ &
$1.0\times10^{8}$ & $10$ & $9.81$ &
$6$ & $174(=29H)$ & $0.061$ & $0.1167$ & $14$ \\
\hline
\end{tabular}
\label{tab:pinn_params}
\end{table}
Figure~\ref{PINN vs Analytical} presents the dispersion curves obtained using the PINN approach alongside the corresponding analytical results with activation function tanh and under 10000 epochs. It can be observed that the PINN predictions closely follow the analytical solution over the entire range of wavenumbers considered. The excellent agreement demonstrates that the proposed physics loss formulation, combined with automatic differentiation and eigenvalue learning of the phase velocity, is capable of accurately capturing the dispersion characteristics of SH-waves in the functionally graded layer–substrate configuration.

\subsection{Error analysis and convergence study}
To evaluate the predictive capability of the proposed PINN framework, a quantitative error analysis is performed by comparing the dimensionless phase velocity predicted by the PINN model with the corresponding analytical solution.
For error quantification, the absolute error is defined as
\begin{equation}
\mathcal{E}(kH) =
\left|
\left(\frac{c}{c_s}\right)_{\text{PINN}}
-
\left(\frac{c}{c_s}\right)_{\text{analytical}}
\right|,
\end{equation}
and the corresponding relative error is computed as
\begin{equation}
\text{Relative Error}
=
\left|
\frac{\left(\frac{c}{c_s}\right)_{\text{PINN}} -
\left(\frac{c}{c_s}\right)_{\text{analytical}}}
{\left(\frac{c}{c_s}\right)_{\text{analytical}}}
\right|.
\end{equation}
For a quantitative evaluation, Table~\ref{PINN_vs_Analytical_quantitative} compares the PINN-predicted and analytical dimensionless phase velocities \(c/c_s\) at selected values of \(kH\), together with the corresponding absolute and relative errors. Over the range \(0.3660 \le kH \le 0.7002\), the relative error varies between \(5.124\times10^{-4}\) and \(2.808\times10^{-2}\). The smallest relative discrepancy occurs at \(kH=0.7002\), whereas the largest is observed at \(kH=0.3660\). Most values remain below approximately \(2.6\times10^{-2}\), with several entries falling below \(10^{-2}\). The absolute error ranges from \(5.359\times10^{-4}\) to \(5.011\times10^{-2}\), remaining predominantly within the order of \(10^{-3}\)–\(10^{-2}\) across the considered wavenumber range. The overall close agreement between the PINN predictions and analytical results confirms the accuracy and robustness of the proposed framework.
\begin{table}[H]
\centering
\caption{Comparison between PINN-predicted and analytical dimensionless phase velocities with absolute and relative errors.}
\label{PINN_vs_Analytical_quantitative}
\renewcommand{\arraystretch}{1.2}
\begin{tabular}{ccccc}
\hline
\textbf{$kH$} & \textbf{PINN $c/c_s$} & \textbf{Analytical $c/c_s$} & \textbf{Abs. Error} & \textbf{Rel. Error} \\
\hline
0.3660 & 1.734044 & 1.784151 & $5.011\times10^{-2}$ & $2.808\times10^{-2}$ \\
0.3917 & 1.647921 & 1.653285 & $5.365\times10^{-3}$ & $3.245\times10^{-3}$ \\
0.4174 & 1.567200 & 1.545226 & $2.197\times10^{-2}$ & $1.422\times10^{-2}$ \\
0.4431 & 1.490476 & 1.455270 & $3.521\times10^{-2}$ & $2.419\times10^{-2}$ \\
0.4688 & 1.416186 & 1.379846 & $3.634\times10^{-2}$ & $2.634\times10^{-2}$ \\
0.4945 & 1.346820 & 1.316202 & $3.062\times10^{-2}$ & $2.326\times10^{-2}$ \\
0.5202 & 1.280742 & 1.262187 & $1.856\times10^{-2}$ & $1.470\times10^{-2}$ \\
0.5460 & 1.217114 & 1.216100 & $1.013\times10^{-3}$ & $8.334\times10^{-4}$ \\
0.5717 & 1.156291 & 1.176586 & $2.029\times10^{-2}$ & $1.725\times10^{-2}$ \\
0.5974 & 1.133165 & 1.142551 & $9.386\times10^{-3}$ & $8.215\times10^{-3}$ \\
0.6231 & 1.110502 & 1.113109 & $2.608\times10^{-3}$ & $2.343\times10^{-3}$ \\
0.6488 & 1.088292 & 1.087538 & $7.535\times10^{-4}$ & $6.928\times10^{-4}$ \\
0.6745 & 1.066526 & 1.065242 & $1.283\times10^{-3}$ & $1.205\times10^{-3}$ \\
0.7002 & 1.045195 & 1.045731 & $5.359\times10^{-4}$ & $5.124\times10^{-4}$ \\
\hline
\end{tabular}
\end{table}
The variation of $\mathcal{E}(kH)$ over the considered wavenumber range is illustrated in Fig.~\ref{fig:absolute_error_combined}. The two-dimensional error distribution (heatmap) is shown in Fig.~\ref{fig:absolute_error_points}, while Fig.~\ref{fig:absolute_error_line} presents the corresponding error profile as a function of $kH$. It is observed that the absolute error remains small and uniformly bounded across the entire computational domain. The absence of irregular oscillatory behavior or localized instabilities indicates that the neural approximation consistently captures the eigenvalue structure of the dispersion relation. Furthermore, the smooth variation of $\mathcal{E}(kH)$ suggests stable optimization and well-behaved eigenvalue convergence during training.

\bfg[htbp]
\centering
\begin{subfigure}[b] {0.45\textwidth}
\includegraphics[width=\textwidth ]{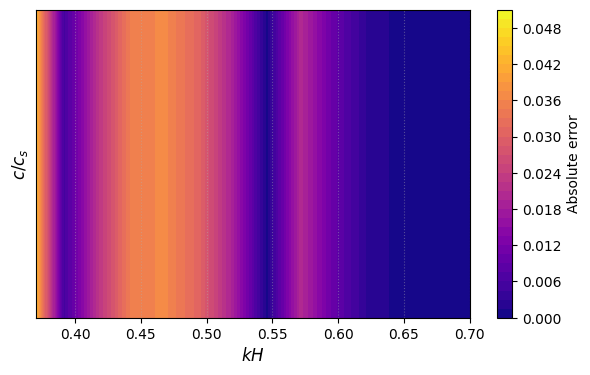}
\caption{}
\label{fig:absolute_error_points}
\end{subfigure}
~
\begin{subfigure}[b] {0.47\textwidth}
\includegraphics[width=\textwidth ]{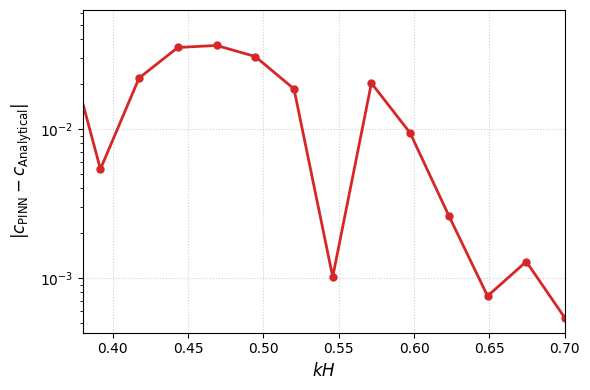}
\caption{}
\label{fig:absolute_error_line}
\end{subfigure}

\begin{figure}
    \centering
    \includegraphics[width=1\linewidth]{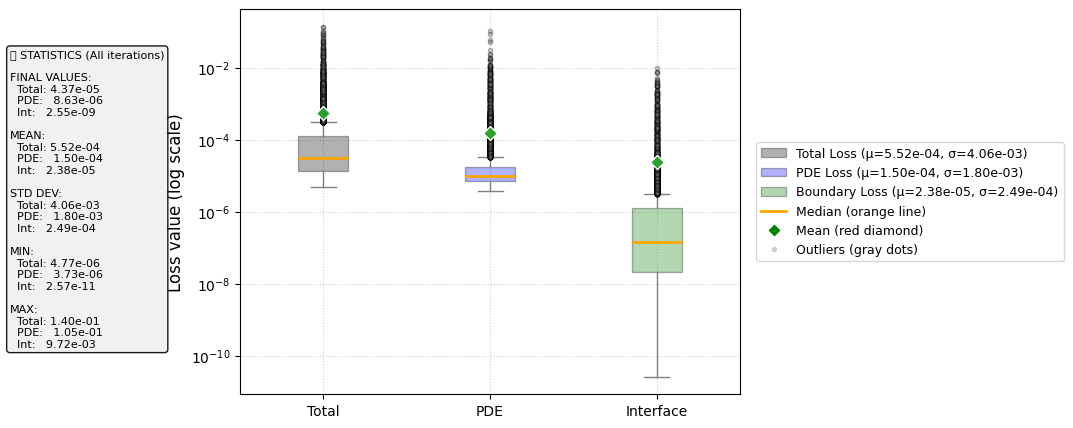}
    \caption{Log-scale boxplot of total, PDE, and interface losses showing PINN convergence.}
    \label{box plot loss history}
\end{figure}

\caption{Absolute error between the PINN-predicted and analytical normalized phase velocities with respect to the dimensionless wavenumber $kH$.}
\label{fig:absolute_error_combined}
\efg
The distribution of the individual loss components over all training iterations is shown in Fig.~\ref{box plot loss history}, where the total, PDE, and interface losses are displayed on a logarithmic scale. The final recorded values are $4.37\times10^{-5}$ (total loss), $8.63\times10^{-6}$ (PDE loss), and $2.55\times10^{-9}$ (interface loss), indicating substantial reduction of all residual components by the end of training.
Statistically, the mean total, PDE, and interface losses are $5.52\times10^{-4}$, $1.50\times10^{-4}$, and $2.38\times10^{-5}$, respectively, with corresponding standard deviations $4.06\times10^{-3}$, $1.80\times10^{-3}$, and $2.49\times10^{-4}$. The medians (orange lines in the box plots) lie below the mean values, reflecting a right-skewed distribution caused by larger loss values during early training iterations. 
Overall, the consistently small magnitudes of the PDE and interface losses demonstrate that both the governing equation residual and the interface continuity conditions are effectively minimized, confirming stable and well-balanced optimization throughout the training process.

\begin{figure}
    \centering
    \includegraphics[width=0.65\linewidth]{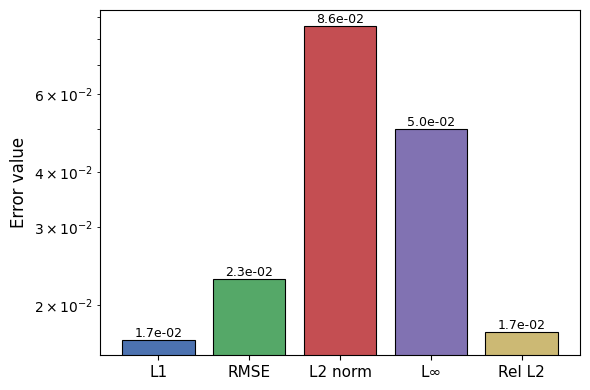}
    \caption{Error analysis of the trained PINN model.}
    \label{comparison of different error norm}
\end{figure}

In addition, a comparative analysis of several error norms is presented in Fig.~\ref{comparison of different error norm}. The evaluation includes the $L_1$ norm (mean absolute error), the $L_2$ norm, the relative $L_2$ norm, the maximum norm ($L_\infty$), and the root mean square error (RMSE), all computed over the discrete set of evaluated wavenumbers.
The computed values are $1.7\times10^{-2}$ for the $L_1$ norm, $8.6\times10^{-2}$ for the $L_2$ norm, $1.7\times10^{-2}$ for the relative $L_2$ norm, $5.0\times10^{-2}$ for the $L_\infty$ norm, and $2.3\times10^{-2}$ for the RMSE. All error measures remain below $10^{-1}$, indicating that the discrepancy between the PINN-predicted and analytical phase velocities is uniformly controlled across the considered wavenumber range.
In particular, the bounded $L_\infty$ norm confirms the absence of significant localized deviations, while the relatively small $L_1$, RMSE, and relative $L_2$ values indicate that both the average and normalized energy of the error are limited. The consistency among these norm-based measures further supports the accuracy and robustness of proposed PINN framework in approximating the dispersion eigenvalue solution.

Table~\ref{tab:activation_comparison} presents a comparative study of the PINN performance obtained with different activation functions, including Sigmoid, Tanh, Swish, Arctan, and Softplus. For each activation, the corresponding analytical expression and resulting relative error are reported.
All activation functions yield nearly identical predictive accuracy, with relative errors clustered tightly around $1.68\times10^{-2}$. Specifically, the relative errors are $1.6789\times10^{-2}$ (Sigmoid), $1.6787\times10^{-2}$ (Tanh), $1.6796\times10^{-2}$ (Swish), $1.6775\times10^{-2}$ (Arctan), and $1.6770\times10^{-2}$ (Softplus). 
Although the Softplus activation produces the smallest relative error, the differences across all tested functions are marginal. These results suggest that, provided sufficient smoothness and differentiability are ensured for higher-order automatic differentiation, the overall performance of the PINN framework remains largely invariant with respect to the specific choice of activation function.
\begin{table}[H]
\centering
\caption{Comparison of PINN performance using different activation functions}
\label{tab:activation_comparison}
\renewcommand{\arraystretch}{1.3}

\begin{tabular}{
| >{\centering\arraybackslash}m{2.3cm}
| >{\centering\arraybackslash}m{4.5cm}
| >{\centering\arraybackslash}m{4.5cm}
| >{\centering\arraybackslash}m{3cm} |
}
\hline
\textbf{Activation Function} &
\textbf{Mathematical Expression} &
\textbf{Activation Function Shape} &
\textbf{Relative Error} \\
\hline

Sigmoid &
$\displaystyle \frac{1}{1+e^{-x}}$ &
\begin{tikzpicture}
\begin{axis}[
width=4cm, height=3cm,
domain=-5:5, samples=100,
ymin=0, ymax=1.1,
xtick={-5,0,5},
ytick={0,0.5,1},
axis lines=left,
tick label style={font=\scriptsize}
]
\addplot[blue, line width=2pt]{1/(1+exp(-x))};
\end{axis}
\end{tikzpicture}
&
${1.6789\times10^{-2}}$ \\
\hline

Tanh &
$\displaystyle \frac{e^{x}-e^{-x}}{e^{x}+e^{-x}}$ &
\begin{tikzpicture}
\begin{axis}[
width=4cm, height=3cm,
domain=-5:5, samples=100,
ymin=-1.1, ymax=1.1,
xtick={-5,0,5},
ytick={-1,0,1},
axis lines=left,
tick label style={font=\scriptsize}
]
\addplot[blue, line width=2pt]{tanh(x)};
\end{axis}
\end{tikzpicture}
&
${1.6787\times10^{-2}}$ \\
\hline

Swish &
$\displaystyle \frac{x}{1+e^{-x}}$ &
\begin{tikzpicture}
\begin{axis}[
width=4cm, height=3cm,
domain=-5:5, samples=100,
ymin=-1, ymax=5,
xtick={-5,0,5},
ytick={0,2,4},
axis lines=left,
tick label style={font=\scriptsize}
]
\addplot [blue, line width=2pt]{x/(1+exp(-x))};
\end{axis}
\end{tikzpicture}
&
$1.6796\times10^{-2}$ \\
\hline

Arctan &
$\displaystyle \arctan(x)$ &
\begin{tikzpicture}
\begin{axis}[
width=4cm, height=3cm,
domain=-5:5, samples=100,
ymin=-1.7, ymax=1.7,
xtick={-5,0,5},
ytick={-1.5,0,1.5},
axis lines=left,
tick label style={font=\scriptsize}
]
\addplot[blue, line width=2pt] {rad(atan(x))};
\end{axis}
\end{tikzpicture}
&
${1.6775\times10^{-2}}$ \\
\hline

Softplus &
$\displaystyle \ln(1+e^{x})$ &
\begin{tikzpicture}
\begin{axis}[
width=4cm, height=3cm,
domain=-5:5, samples=100,
ymin=0, ymax=6,
xtick={-5,0,5},
ytick={0,2,4},
axis lines=left,
tick label style={font=\scriptsize}
]
\addplot [blue, line width=2pt]{ln(1+e^x)};
\end{axis}
\end{tikzpicture}
&
$\mathbf{1.6770\times10^{-2}}$ \\
\hline

\end{tabular}
\end{table}

\begin{figure}
    \centering
    \includegraphics[width=0.7\linewidth]{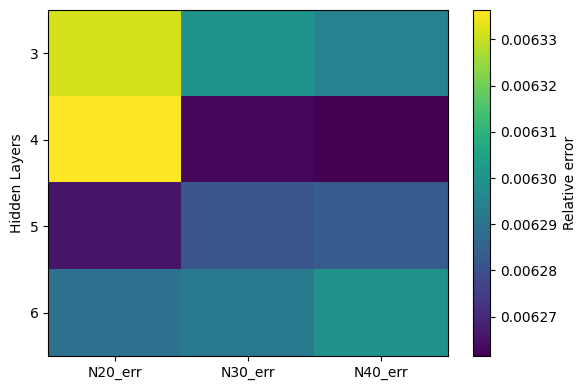}
    \caption{Heat map showing the relative error in the PINN-predicted normalized phase velocity $c/c_S$ for various neural network architectures with different hidden-layer depths and neuron counts, evaluated at $kH = 0.7$.}

    \label{neural network architecture}
\end{figure}

Table~\ref{tab:pinn_arch} summarizes the performance of the model for networks consisting of 3-6 hidden layers with 20, 30, and 40 neurons per layer. The analytical solution remains fixed at $c/c_S=1.045731$ at $kH=0.7$, while the PINN predictions remain very close to this value for all tested configurations. The relative error is observed to be of the order $10^{-3}$ in all cases, indicating the strong predictive capability of the proposed PINNs formulation. 
In the present simulations, the learning rate is set to $5\times10^{-5}$, the hyperbolic tangent ($\tanh$) function is employed for activation, and the Adam optimizer is utilized for training the neural network.
It is observed that increasing the number of neurons slightly improves the accuracy for shallower networks, whereas increasing the depth beyond a moderate number of layers does not produce a significant improvement in prediction accuracy.
Figure~\ref{neural network architecture} presents the heat map corresponding to different numbers of hidden layers and neurons, along with the associated relative errors. The graphical illustration confirms that the prediction error remains uniformly small across the tested architectures, demonstrating the stability and convergence of the PINNs model.
\begin{table}[H]
\centering
\caption{Performance of the present PINNs model using different network architectures for $kH=0.7$.}
\label{tab:pinn_arch}
\renewcommand{\arraystretch}{1.3}
\begin{tabular}{|ccccc|}
\hline
\textbf{Hidden layers} & \textbf{Neurons} & \textbf{Analytical $c/c_S$} & \textbf{PINNs $c/c_S$} & \textbf{Relative errors} \\
\hline
\multirow{3}{*}{3} 
& 20 &  & 1.039110 & $6.332\times10^{-3}$ \\
& 30 &  1.045731  &1.039143 & $6.300\times10^{-3}$ \\
& 40 &   & 1.039148 & $6.295\times10^{-3}$ \\
\hline

\multirow{3}{*}{4} 
& 20 &   & 1.039105 & $6.336\times10^{-3}$ \\
& 30 &  1.045731 & 1.039182 & $6.263\times10^{-3}$ \\
& 40 &   & 1.039183 & $6.261\times10^{-3}$ \\
\hline

\multirow{3}{*}{5} 
& 20 &  & 1.039179 & $6.266\times10^{-3}$ \\
& 30 & 1.045731  & 1.039162 & $6.281\times10^{-3}$ \\
& 40 &  & 1.039161 & $6.283\times10^{-3}$ \\
\hline

\multirow{3}{*}{6} 
& 20 &  & 1.039154 & $6.290\times10^{-3}$ \\
& 30 &  1.045731 & 1.039151 & $6.292\times10^{-3}$ \\
& 40 &  & 1.039143 & $6.300\times10^{-3}$ \\
\hline
\end{tabular}
\end{table}

\section{Conclusion}
\label{Conclusion}
This study presents a physics-informed neural network (PINN) framework for analyzing dispersion characteristics of SH-waves in an initially stressed magnetoelastic orthotropic functionally graded layer lying on an initially stressed functionally graded orthotropic stratum under the influence of gravity. 
The proposed PINN framework integrates the governing differential equations together with the boundary constraints into the learning procedure, enabling efficient computation of the dispersion relation without employing traditional mesh-based numerical discretization.
The accuracy and robustness of the proposed framework are evaluated through comprehensive error analysis using the $L_1$, $L_2$, and $L_\infty$ norms, along with relative and absolute error measures. In addition, the influence of different neural network architectures, obtained by varying the number of hidden layers and neurons, is investigated. The influence of distinct activation functions on predictive performance of the PINN framework is also investigated.
The principal findings of the present work are outlined as follows:
\begin{itemize}
\item The heterogeneity parameters significantly influence SH-wave propagation in the graded media. Increasing the heterogeneity parameter of the upper layer reduces the phase velocity, indicating a decrease in the effective stiffness of the medium. Conversely, a higher value of heterogeneity parameter in the lower half-space results in an increase in the phase velocity, suggesting enhanced effective rigidity and allowing quicker SH-wave propagation.
\item The initial stress influences the propagation of SH waves in the graded structure. In the upper layer, an increase in the initial stress leads to a slight increase in the phase velocity, indicating that the presence of initial stress enhances the effective stiffness of the medium. In contrast, in the underlying half-space, a decrease in the initial stress results in a reduction in the phase velocity, showing that lower initial stress weakens the effective rigidity and slows down the propagation of SH waves. The gravity parameter in Biot’s formulation also lowers the phase velocity as its value increases. In contrast, an increase in layer thickness enhances the phase velocity, suggesting that a thicker graded layer supports faster SH-wave propagation.
\item Magnetic field parameters influence the propagation of SH waves. An increase in the magnetic field angle results in a slight rise in the phase velocity, indicating an enhancement in the effective stiffness of the medium. In contrast, higher induced magnetic permeability modifies the magnetoelastic coupling and slightly reduces the phase velocity.
\item The dispersion curves obtained using the PINN framework show excellent agreement with the analytically derived results across the entire wavenumber range, demonstrating that the proposed physics-informed loss formulation combined with automatic differentiation and eigenvalue learning can accurately capture the dispersion behavior of SH-waves in the considered functionally graded layer–half-space system.
\item A quantitative comparison between the PINN predictions and analytical phase velocities demonstrates high accuracy of the proposed framework, with relative errors ranging from $5.124\times10^{-4}$ to $2.808\times10^{-2}$ and absolute errors remaining predominantly within the order of $10^{-3}$–$10^{-2}$ across the considered wavenumber range, confirming the robustness and reliability of the PINN-based approach.
\item The distribution of the absolute error $\mathcal{E}(kH)$ remains small and uniformly bounded across the considered wavenumber range, with no irregular oscillations or localized instabilities. This behavior indicates stable optimization during training and confirms that the PINN model successfully captures the eigenvalue structure of the dispersion relation.
\item The training evolution of the loss terms indicates that the total, PDE, and interface losses decrease progressively, reaching final values of $4.37\times10^{-5}$, $8.63\times10^{-6}$, and $2.55\times10^{-9}$, respectively. This indicates effective minimization of the governing equation residuals and interface continuity conditions, confirming stable and well-balanced optimization of the PINN 
\item A comparative analysis of multiple error norms, including $L_1$, $L_2$, relative $L_2$, $L_\infty$, and RMSE, shows that all error measures remain below $10^{-1}$ across the considered wavenumber range. The bounded $L_\infty$ norm indicates the absence of localized deviations, while the small $L_1$, relative $L_2$, and RMSE values confirm that both the average and normalized energy of the error remain limited, demonstrating the accuracy and robustness of proposed PINN framework.
\item The investigation of different network architectures shows that the PINN predictions remain in close agreement with the analytical phase velocity for all tested configurations. The relative error remains of the order $10^{-3}$, demonstrating the strong predictive capability of the proposed framework. It is also observed that increasing the number of neurons slightly improves the accuracy for shallow networks, while increasing the depth beyond a moderate number of layers does not lead to significant improvement in prediction accuracy.
\end{itemize}
This study has important applications in engineering and biomedical systems involving wave propagation in heterogeneous layered materials. It highlights the influence of material grading, initial stress, magnetic field effects, and gravity on SH-wave dispersion, while demonstrating the effectiveness of PINNs for accurate dispersion prediction in complex graded media.

\noindent {\bf Acknowledgements}\\
The authors express their sincere gratitude to the National Institute of Technology Hamirpur for providing research facilities to Ms.~Diksha during her doctoral studies. The authors also acknowledge the University Grants Commission (UGC) for providing the research fellowship.\\
\noindent {\bf Conflicts of interest}\\
The authors declare no conflict of interest.\\
\bibliographystyle{unsrt}
\bibliography{refrences}
\end{document}